%% file: B2G-24-008_temp.tex
\documentclass[11pt,twoside,a4paper,cmspaper,final,collab]{cms-tdr}
\def\svnVersion{18ea797-D}\def\svnDate{2026/09/28}\def\cmsCernNoTag{CERN-EP-2025-297}\def\cmsCernDate{\today}\def\cmsMessage{Published in the Journal of High Energy Physics as \href{http://dx.doi.org/10.1007/JHEP10(2026)013}{\doi{10.1007/JHEP10(2026)013}.}}
\begin{document}\cmsNoteHeader{B2G-24-008}

\newcommand{\PA}{\HepParticle{A}{}{}\xspace}
\newcommand{\Hp}{\ensuremath{\HepParticle{\PH}{}{+}\xspace}}
\newcommand{\mHpm}{\ensuremath{m_{\Hpm}}\xspace}
\newcommand{\mt}{\ensuremath{m_{\PQt}}\xspace}
\newcommand{\wz}{\ensuremath{\PW\PZ}\xspace}
\newcommand{\ww}{\ensuremath{\PW\PW}\xspace}
\newcommand{\ttW}{\ensuremath{\ttbar\PW}\xspace}
\newcommand{\ttZ}{\ensuremath{\ttbar\PZ}\xspace}
\newcommand{\ttH}{\ensuremath{\ttbar\Ph}\xspace}
\newcommand{\ttV}{\ensuremath{\ttbar\PV}\xspace}
\newcommand{\zjet}{\ensuremath{\PZ}+\text{jets}\xspace}
\newcommand{\wjet}{\ensuremath{\PW}+\text{jets}\xspace}
\newcommand{\ttX}{\ensuremath{\ttbar\text{X}}\xspace}
\newcommand{\ttXY}{\ensuremath{\ttbar\text{XY}}\xspace}
\newcommand{\ttHH}{\ensuremath{\ttbar\Ph\Ph}\xspace}
\newcommand{\ttWH}{\ensuremath{\ttbar\PW\Ph}\xspace}
\newcommand{\ttWW}{\ensuremath{\ttbar\PW\PW}\xspace}
\newcommand{\ttWZ}{\ensuremath{\ttbar\PW\PZ}\xspace}
\newcommand{\ttZH}{\ensuremath{\ttbar\PZ\Ph}\xspace}
\newcommand{\ttZZ}{\ensuremath{\ttbar\PZ\PZ}\xspace}
\newcommand{\ttVH}{\ensuremath{\ttbar\PV\Ph}\xspace}
\newcommand{\ttVV}{\ensuremath{\ttbar\PV\PV}\xspace}
\newcommand{\tZ}{\ensuremath{\PQt\PZ}\xspace}
\newcommand{\tW}{\ensuremath{\PQt\PW}\xspace}
\newcommand{\tttj}{\ensuremath{\ttbar\PQt\text{j}}\xspace}
\newcommand{\tttW}{\ensuremath{\ttbar\PQt\PW}\xspace}
\newcommand{\tttt}{\ensuremath{\ttbar\ttbar}\xspace}
\newcommand{\VVV}{\ensuremath{\PV\PV\PV}\xspace}
\newcommand{\pp}{\ensuremath{\Pp\Pp}\xspace}
\newcommand{\pb}{\unit{pb}}
\newcommand{\pptoqHpm}{\ensuremath{\pp\to(\PQq)\Hpm}\xspace}
\newcommand{\qqh}{\ensuremath{\pptoqHpm\to(\PQq)\PQt\PQb}\xspace}
\newcommand{\pptobHpmwoPar}{\ensuremath{\pp\to\PQb\Hpm}\xspace}
\newcommand{\pptobHpm}{\ensuremath{\pp\to(\PQb)\Hpm}\xspace}
\newcommand{\pptobHp}{\ensuremath{\pp\to\PQb\Hp}\xspace}
\newcommand{\pptocHpm}{\ensuremath{\pp\to\PQc\Hpm}\xspace}
\newcommand{\qqhonlyplus}{\ensuremath{\pp\to\Hp}\xspace}
\newcommand{\qqhplusminus}{\ensuremath{\pp\to\Hpm}\xspace}
\newcommand{\rhotu}{\ensuremath{\rho_{\PQt\PQu}}\xspace}
\newcommand{\rhotc}{\ensuremath{\rho_{\PQt\PQc}}\xspace}
\newcommand{\rhott}{\ensuremath{\rho_{\PQt\PQt}}\xspace}
\newcommand{\mur}{\ensuremath{\mu_\text{r}}\xspace}
\newcommand{\muf}{\ensuremath{\mu_\text{f}}\xspace}
\newcommand{\Imini}{\ensuremath{I_{\text{mini}}}\xspace}
\newcommand{\SIP}{\ensuremath{\mathrm{SIP}_\mathrm{3D}}\xspace}
\newcommand{\lpt}{\ensuremath{\pt^{\Pell}}\xspace}
\newcommand{\mW}{\ensuremath{m_{\PW}}\xspace}
\newcommand{\pDNNtwob}{\ensuremath{\text{pDNN}_{2\PQb}}\xspace}
\newcommand{\pDNNthreeb}{\ensuremath{\text{pDNN}_{3\PQb}}\xspace}
\newcommand{\crosssectionbH}{\ensuremath{\sigma(\pp\to\PQb\PH^\pm)}\xspace}
\newcommand{\crosssectionbHwPar}{\ensuremath{\sigma(\pp\to(\PQb)\PH^\pm)}\xspace}
\newcommand{\crosssection}{\ensuremath{\sigma({\pp\to(\PQq)\PH^\pm})}\xspace}
\newcommand{\crosssectionlight}{\ensuremath{\sigma(\pp\to\PH^\pm)}\xspace}
\newcommand{\htb}{\ensuremath{\PH^\pm\to\PQt\PQb}\xspace}
\newcommand{\tblnu}{\ensuremath{\PQt\to\PQb \Pell\PGn}\xspace}
\newcommand{\brhtb}{\ensuremath{\mathcal{B}(\htb, \tblnu)}\xspace}
\newcommand{\toppt}{\ensuremath{\pt^\PQt}\xspace}
\newcommand{\pznu}{\ensuremath{p_{z}^{\PGn}}\xspace}
\providecommand{\DEEPJET}{\textsc{DeepJet}\xspace}
\providecommand{\cmsTable}[1]{\resizebox{\textwidth}{!}{#1}}
\newlength\cmsTabSkip\setlength{\cmsTabSkip}{2ex}

\cmsNoteHeader{B2G-24-008}
\title{Search for charged Higgs bosons decaying into top and bottom quarks in lepton+jets final states in proton-proton collisions at \texorpdfstring{$\sqrt{s} = 13\TeV$}{sqrt(s) =13 TeV}}

\author[cern]{CMS Collaboration}

\date{\today}

\abstract{
A search is presented for charged Higgs bosons (\Hpm) in proton-proton (\pp) collision events via the \pptobHpm processes, with \Hpm decaying into top (\PQt) and bottom (\PQb) quarks. 
The search targets final states with one lepton, missing transverse momentum, and two or more \PQb jets. 
The analysis is based on data collected at a center-of-mass energy of 13\TeV with the CMS detector at the CERN LHC, corresponding to an integrated luminosity of 138\fbinv. We search for charged Higgs bosons in the 200\GeV to 1\TeV mass range. 
The results are interpreted within the generalized two-Higgs-doublet model (g2HDM). 
This model predicts additional Yukawa couplings of the Higgs bosons to the top quark $\rhott$, the top and charm quark $\rhotc$, and the top and up quark $\rhotu$. This search focuses on the real components of $\rhott$ and $\rhotc$, which are probed up to values of unity. 
An excess is observed with respect to the standard model expectation with a local significance of 2.4 standard deviations for a signal with an \Hpm boson mass (\mHpm) of 600\GeV. 
Limits are derived on the product of the cross section \crosssectionbHwPar and branching fraction \brhtb, where $\Pell=\Pe, \PGm$. 
The values of $\rhotc \gtrsim$ 0.11--0.38 are excluded at 95\% confidence level, depending on the $\mHpm$ and \rhott assumptions. 
The results represent the first search for charged Higgs bosons within the g2HDM framework in the single-lepton final states and complement the existing results on additional neutral Higgs bosons perfomed 
in the same-sign top quark pair, three top quark, or four top quark final states, constraining extra Yukawa couplings for the neutral Higgs bosons below the \TeV scale. 
}

\hypersetup{
pdfauthor={CMS Collaboration},%
pdftitle={Search for charged Higgs bosons decaying into top and bottom quarks in lepton+jets final states in proton-proton collisions at sqrt(s) = 13 TeV},
pdfsubject={CMS},
pdfkeywords={CMS, Higgs, beyond standard model}}

\maketitle 

\section{Introduction}
In 2012, the ATLAS and CMS experiments at CERN discovered a scalar boson with a mass of 125\GeV~\cite{Aad:2012tfa,Chatrchyan:2012ufa,CMS:2013btf}.  
The properties of this boson have been scrutinized in great detail and found consistent with those predicted for the Higgs boson in the standard model (SM)~\cite{CMS:2022dwd,ATLAS:2022vkf}.
The CERN LHC Run 2 (2015--2018) data enabled more precise measurements of the Higgs boson properties and put further constraints on possible contributions from new physics beyond the SM.
The ATLAS and CMS experiments have measured the Yukawa couplings of the 125\GeV Higgs boson to the top quark (\PQt)~\cite{Sirunyan:2018hoz,ATLAS:2018mme,Sirunyan:2020eds}, 
bottom quark (\PQb)~\cite{Sirunyan:2018kst,ATLAS:2018kot}, tau lepton (\PGt)~\cite{ATLAS:2015xst,Sirunyan:2017khh}, and the muon (\PGm)~\cite{CMS:2020xwi,ATLAS:2025coj}, 
as well as constrained the couplings to charm quarks (\PQc)~\cite{ATLAS:2024yzu,CMS:2025dsh}.
Although the measured couplings in all cases are found to be consistent with the SM expectations within uncertainties, 
having established the existence of one scalar doublet, it is natural to question if a second doublet exists.

A second doublet is predicted in the two-Higgs-doublet model (2HDM) framework, implying the existence of two charged Higgs bosons (\Hpm), a new scalar boson (\PH), 
and a pseudoscalar boson (\PA), in addition to the SM-like Higgs boson (\Ph).
There are several main classes of 2HDM extensions~\cite{Branco:2011iw}:
\begin{itemize}
  \item \textit{2HDM-I}: All quarks and charged leptons couple to the same Higgs doublet.
  \item \textit{2HDM-II}: The \PQu-type quarks couple to one doublet, while \PQd-type quarks and charged leptons couple to the other.
  \item \textit{2HDM-X}: All quarks couple to one doublet, and charged leptons couple to the other.
  \item \textit{2HDM-Y}: This is the flipped version of the 2HDM-X.
  \item \textit{2HDM-III} or \textit{generalized 2HDM (g2HDM)}: Extra Yukawa couplings are present, and unlike the other types listed above, no $\mathbb{Z}_2$ symmetry is imposed.
\end{itemize}
Earlier LHC searches for \Hpm performed in the context of the 2HDM-II, including Run~2 results, are presented in Refs.~\cite{CMS:2019bfg,CMS:2020imj,ATLAS:2021jol,ATLAS:2021upq,CMS:2021wlt,CMS:2022jqc,ATLAS:2022zuc,ATLAS:2022pbd,ATLAS:2023bzb,ATLAS:2024oqu,ATLAS:2024txt,ATLAS:2024hya,ATLAS:2024rcu}.

In the g2HDM, the absence of a $\mathbb{Z}_2$ symmetry allows for flavor-changing neutral current (FCNC) interactions involving the \PA and \PH bosons, while such interactions remain suppressed for the \Ph boson.
In the g2HDM, the couplings of the spin-0 particles \Ph, \PH, \PA, and \Hpm to fermions are regulated by the Yukawa couplings, 
the mixing angle ($\gamma$) between the \Ph and \PH scalar particles, and the elements of the Cabibbo--Kobayashi--Maskawa (CKM) matrix~\cite{Davidson:2005cw}.
The largest contributions to the \PQc{}\PQb{}\Hpm and \PQt{}\PQb{}\Hpm couplings come from $\rhotc V_\mathrm{tb}$ and $\rhott V_\mathrm{tb}$, respectively. 
Unlike in the 2HDM-II, where the corresponding couplings are CKM-suppressed, 
here, the $\rhotc V_\mathrm{tb}$ and $\rhott V_\mathrm{tb}$ terms appear without suppression.
As a result, the \Hpm production cross section in the g2HDM can be significantly larger 
than in the 2HDM-II. 
A more detailed discussion can be found in Ref.~\cite{Ghosh:2019exx}.

With the additional Yukawa couplings, many parameters and extra processes become possible, but  
there are tight constraints from electroweak precision measurements, flavor physics, and direct searches at the LHC. 
However, extra Higgs bosons at mass scales ${<}1\TeV$ at the LHC 
could still be hidden by the fermion mass-mixing hierarchy~\cite{Hou:1991un} or the alignment mechanisms~\cite{Hou:2017hiw}, where
the mixing angle $\gamma$ between the two $CP$-even scalars is close to $\pi/2$, \ie,
$\cos\gamma\approx0$. 
Therefore, the lack of a hint of FCNC interactions of the SM Higgs boson at the LHC may be a result of the amplitude suppression by the alignment mechanism. 
The extra Yukawa couplings of the additional Higgs bosons to the top quark $\rhott$, to the top and charm quark $\rhotc$, 
and to the top and up quark $\rhotu$ can assume large values~\cite{Hou:2017hiw}, and combined with 
the Higgs quartic couplings of order 1, they can also explain electroweak baryogenesis~\cite{Ghosh:2019exx,Fuyuto:2017ewj,Fuyuto:2019svr,Kohda:2017fkn}.
 These arguments strongly motivate direct experimental searches for extra Yukawa couplings. 
 
\begin{figure}[ht]
   \centering
   \includegraphics[width=0.32\textwidth]{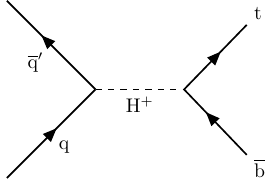}\hfill
   \includegraphics[width=0.31\textwidth]{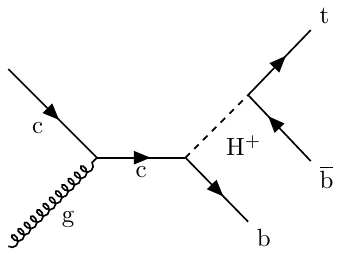}\hfill
   \includegraphics[width=0.32\textwidth]{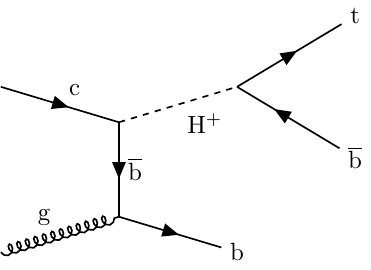}
   \caption{Representative tree-level Feynman diagrams for the signal processes: $\qqhonlyplus,~\PH^+\to\PQt\PAQb$ (left), and $\pptobHp,\PH^+\to\PQt\PAQb$ (middle and right).}
   \label{fig:Hp}
 \end{figure}

Previous searches for extra Yukawa couplings based on the g2HDM were focused on \PA and \PH bosons  
and no significant excess above the background prediction was observed in either case, leading to 
strong bounds on \rhotu~\cite{ATLAS:2023tlp,CMS:2024jcz}. 
In this paper, we focus on \Hpm and further explore the extra Yukawa couplings \rhotc and \rhott, while setting \rhotu to zero. 
We consider \PHpm production in proton-proton (\pp) collisions in the \qqh channels, with \PQq being a \PQb or \PQc quark. 
This process is dominated by the \qqhplusminus and \pptobHpmwoPar contributions.
The \pptocHpm is suppressed by the parton distribution function (PDF) of the bottom quark in the proton.
This process is taken into account for completeness and is categorized together with \qqhplusminus.
Representative Feynman diagrams for the signal processes are displayed in Fig.~\ref{fig:Hp}.
Considering the decay \htb and top quark decays with one isolated electron or muon in the final state, 
the signature is expected to contain two or three \PQb quarks at tree level, as well as a lepton and  
missing transverse momentum (\ptmiss).  
In the 2HDM-II, \Hpm production via these processes is suppressed,  
resulting in negligible cross sections and limited experimental sensitivity. 
Consequently, such channels have not been the focus of previous experimental searches. 
In contrast, in the g2HDM, they become dominant because of the presence of an additional Yukawa coupling \rhotc.
These processes result in low jet multiplicity events, which are enhanced in the g2HDM but not in the 2HDM-II. 
The $\PQq\Pg\to\PQt\Hpm$ processes, permitted in both the 2HDM-II and g2HDM, are omitted from this study because their
cross sections are at least an order of magnitude smaller (for $\rhott\sim\rhotc\sim1$) than those of the processes considered in this search.

The paper is organized as follows: Section~\ref{sec:CMSdet} describes the CMS detector. 
In Section~\ref{sec:sim}, descriptions of the data and the simulation samples used in the analysis are given. 
The event selection and reconstruction procedure of the physics objects are given in 
Sections~\ref{sec:eventrec} and ~\ref{sec:eventsel}, respectively.
Background estimation from control samples in the data is discussed in Section~\ref{sec:bkgd}, and 
the signal extraction procedure in Section~\ref{sec:signalextract}. 
Section~\ref{sec:systematics} discusses the sources of systematic uncertainty.
The results of the analysis are presented in Section~\ref{sec:results}. 
Finally, a summary of the paper is given in Section~\ref{sec:summary}.
Tabulated results are accessible from the HEPData record for this analysis~\cite{hepdata}.

\section{The CMS detector}
\label{sec:CMSdet}
 The CMS apparatus~\cite{CMS:2008xjf,CMS:2023gfb} is a multipurpose, nearly hermetic detector, designed to trigger on~\cite{CMS:2020cmk,CMS:2016ngn,CMS:2024aqx} and 
 identify electrons, muons, photons, and (charged and neutral) hadrons~\cite{CMS:2018rym,CMS:2020uim,CMS:2014pgm}. 
 Its central feature is a superconducting solenoid of 6\unit{m} internal diameter, providing a magnetic field of 3.8\unit{T}. 
 Within the solenoid volume are a silicon pixel and strip tracker, a lead tungstate crystal electromagnetic calorimeter (ECAL), 
 and a brass and scintillator hadron calorimeter, each composed of a barrel and two endcap sections. 
 Forward calorimeters extend the pseudorapidity ($\eta$) coverage provided by the barrel and endcap detectors. 
 Muons are reconstructed using gas-ionization detectors embedded in the steel flux-return yoke outside the solenoid. 
 More detailed descriptions of the CMS detector, together with a definition of the coordinate system used and the relevant 
 kinematic variables, can be found in Refs.~\cite{CMS:2008xjf,CMS:2023gfb}. 

Events of interest are selected using a two-tiered trigger system. The first level, composed of custom hardware processors, 
uses information from the calorimeters and muon detectors to select events at a rate of around 100\unit{kHz} within a fixed latency 
of about 4\mus~\cite{CMS:2020cmk}.
The second level, known as the high-level trigger, consists of a farm of processors running a version of the full event reconstruction 
software optimized for fast processing, and reduces the event rate to around 1\unit{kHz} before data storage~\cite{CMS:2016ngn,CMS:2024aqx}. 

 \section{Data and simulated event samples}
 \label{sec:sim}
The data were collected with the CMS detector at the LHC during the years 2016 to 2018, and correspond to an integrated luminosity of 138\fbinv.
Data are selected online with a combination of single-muon, single-electron, and single-photon triggers to maximize signal efficiency. 
In particular, single-photon triggers are used to recover the efficiency for high-energy electrons.

The signal and background processes are simulated using several Monte Carlo (MC) generators. 
The $(\PQq)\PHpm$ signal samples are simulated at leading order (LO) in perturbative quantum chromodynamics (QCD), 
with up to two additional noncollinear high-transverse momentum (\pt) partons in the matrix element calculations using  \MGvATNLO v2.6.5 ~\cite{Alwall:2014hca}, 
and the hard scattering simulation is interfaced with parton shower jets using the MLM~\cite{Alwall:2007fs} matching prescription.
The signal samples are generated considering all leptonic final states coming from the top quark decay chain.
Several signal samples are simulated assuming nonzero values of $\rhott$ and $\rhotc$ selected from 0.1, 0.4, 0.6, and 1, 
and assuming an \PHpm boson mass (\mHpm)  between 200 and 1000\GeV. 
Nine \PHpm boson mass points are considered from 200 to 1000\GeV in 100\GeV steps, chosen based on the mass resolution, 
and one more point at 350\GeV is added to improve the sensitivity around the \ttbar threshold.
In this analysis, \PHpm decays to \PH or \PA are not considered.

Background events generated by \MGvATNLO at next-to-LO (NLO) in perturbative QCD (pQCD) are 
\ttW, \tZ, \VVV, $s$-channel single top quark  production, and \zjet processes, and the ones generated at LO precision are  
\ttZ, \wjet, \ttVV, \ttVH, \ttHH, \tttj, \tttW, \tttt, and multijet production.
Here, \PV represents a \PW or \PZ boson.
Multijet events refer to the events composed uniquely of jets produced through the QCD interaction.  
Among these processes, FxFx matching~\cite{Frederix:2012ps} is used for \ttW and \zjet, while MLM matching~\cite{Alwall:2007fs} is used for \wjet. 
The \POWHEG~v2.0 generator~\cite{Nason:2004rx,Frixione:2007vw,Alioli:2010xd} is used to simulate events at NLO  in pQCD for \ttbar~\cite{Frixione:2007nw},
$\PWpm\PWmp$~\cite{Melia:2011tj}, single top quark production in association with a \PW boson (\tW)~\cite{Re:2010bp}, 
single top quark $t$-channel production~\cite{Alioli:2009je}, and \ttH production~\cite{Hartanto:2015uka}.  
The \PYTHIA~v8.240~\cite{Sjostrand:2014zea} generator is used to simulate \wz and \ww events at LO. 
If a higher-order cross section calculation exists, the yields of backgrounds are normalized to the corresponding value, as discussed below. 

The predicted \ttbar production cross section is 834$^{+38}_{-43}$ pb 
as calculated with the \textsc{Top++} v2.0 program at next-to-NLO (NNLO) in pQCD, 
including soft-gluon resummation to next-to-next-to-leading logarithmic (NNLL) order 
(as discussed in Ref.~\cite{Czakon:2011xx} and references therein), 
and assuming a top quark mass $\mt=172.5\GeV$. 
The predicted cross sections of single top quark and single top antiquark production in the {\it t} channel 
are $\sigma(\PQt\PQq)=134.2^{+1.9}_{-1.3}$\pb and $\sigma(\PAQt\PQq)=80.0^{+2.3}_{-1.8}$\pb, respectively. 
The cross sections are calculated with the \MCFM program~\cite{Campbell:2020fhf} at NNLO in pQCD assuming $\mt=172.5\GeV$. 
For both \ttbar and single top quark cross sections quoted above, the uncertainties 
include the ones due to the choice of the renormalization and factorization scales, \mur and 
\muf, the uncertainties in the PDFs and \alpS. 
The inclusive \PQt{}\PW production cross section is calculated at NLO in pQCD with the addition of third-order corrections of soft-gluon emissions by 
resumming NNLL terms~\cite{Kidonakis:2021vob}. 
For single top quark $s$-channel production, the cross section at NLO in pQCD is calculated using the \textsc{Hathor} v2.1 program~\cite{Kant:2014oha,Aliev:2010zk}.

The cross sections for \wjet events are normalized to NNLO calculations using the \textsc{FEWZ} v3.1 program~\cite{Li:2012wna}.
The inclusive cross section for \ttZ, calculated at NLO+NNLL, is taken from 
Ref.~\cite{Kulesza:2018tqz}.
Cross sections at NLO are used for \ttW\cite{Frederix:2021agh}, \ttH~\cite{LHCHiggsCrossSectionWorkingGroup:2016ypw}, 
\ttHH~\cite{Frederix:2014hta}, \ttVH~\cite{LHCHiggsCrossSectionWorkingGroup:2016ypw}, \ttVV~\cite{Maltoni:2015ena}, 
\tZ~\cite{CMS:2017wrv}, and \tttt~\cite{vanBeekveld:2022hty} processes.

Simulated samples are interfaced with \PYTHIA~v8.240 for parton showering, fragmentation, and hadronization.
The underlying event is also modeled with \PYTHIA~v8, using the CP5 tune~\cite{Sirunyan:2019dfx}.
The PDFs are taken from the NNPDF3.1 NNLO set~\cite{NNPDF:2017mvq}.
Finally, for simulated event samples, the detector response is modeled with the \GEANTfour package \cite{Agostinelli:2002hh}.

The simulated events include effects of additional \pp interactions within the same or nearby bunch crossings (pileup) 
by overlaying simulated minimum bias collisions on the hard scattering interaction, 
with the multiplicity matched to that observed in data~\cite{CMS:2020ebo}.
To match the measured top quark \pt~\cite{CMS:2021vhb} distribution, the simulated \ttbar events are reweighted 
according to the NNLO-to-NLO $k$-factor, by multiplying the \pt 
of the top quarks at the generator level (\toppt) by the function $\exp(0.0615-0.0005\toppt/\GeV)$. 

\section{Event reconstruction}
\label{sec:eventrec}
{\tolerance=400
A global ``particle-flow" (PF) algorithm~\cite{CMS:2017yfk} aims to reconstruct all individual particles in an event (PF candidates), combining information provided by the inner tracker and the calorimeters, 
with data from the muon detectors. 

The primary vertex (PV) is taken to be the vertex corresponding to the hardest scattering in the event, evaluated using tracking information alone, as described in Section~9.4.1 of Ref.~\cite{CMS-TDR-15-02}. 

Muons are reconstructed within $\abs{\eta}<2.4$ via a combination of the measurements with the tracker and the muon spectrometers 
in a global fit~\cite{CMS:2018rym}. 
Electrons are reconstructed combining the tracker and ECAL measurements~\cite{CMS:2020uim} within $\abs{\eta}<2.5$.
Electrons that fall in the region between the ECAL barrel and endcaps ($1.442<\abs{\eta}<1.556$) are discarded. 

Jets are built from PF candidates using the anti-\kt algorithm~\cite{Cacciari:2008gp,Cacciari:2011ma} with a distance parameter of 0.4.
Pileup interactions can contribute additional tracks and calorimetric energy depositions, increasing the apparent jet momentum. 
To mitigate this effect, tracks identified to be originating from pileup are discarded, and an offset correction is applied to correct for remaining contributions~\cite{CMS:2020ebo}. Jet energy corrections are derived from simulation studies so that the average measured energy of jets becomes identical to that of particle-level jets.
In situ measurements of the momentum balance in dijet, photon+jet, \PZ{}+jets, and multijet events are used to determine any residual differences between the jet energy scale in data and in simulation, and appropriate corrections are made ~\cite{CMS:2016lmd}.
The energy of each jet is smeared by an amount depending on its \pt and $\eta$ to adjust the jet energy resolution in simulation to match the one in data.
Additional selection criteria are applied to each jet to remove jets potentially dominated by instrumental effects or reconstruction failures~\cite{CMS:2020ebo}. 

The vector \ptvecmiss is computed as the negative vector sum of the \pt of all the PF candidates in an event, and its magnitude is denoted as \ptmiss~\cite{CMS:2019ctu}. The \ptvecmiss is modified to account for corrections to the energy scale of the reconstructed jets in the event. 

Hadronic \PGt lepton decays (\tauh) are reconstructed from jets, using the hadrons-plus-strips algorithm~\cite{CMS:2018jrd}, 
 which combines one or three tracks with energy deposits in the calorimeters, to identify the \PGt decay modes.
The \textsc{DeepTau} algorithm~\cite{CMS:2022prd} is used to distinguish genuine \tauh decays from jets originating from electrons, muons,
or the hadronization of quarks or gluons.
The chosen working point provides a \tauh identification efficiency of 
approximately 90\% and a misidentification rate of about 5\% for \tauh candidates with \pt up to 200\GeV~\cite{CMS-DP-2019-033,CMS:2022prd}.
The events that contain \tauh candidates with $\pt>20\GeV$ and $\abs{\eta}<2.3$ are vetoed. 
\par}

\section{Event selection}
\label{sec:eventsel}
{\tolerance=400
Events are selected to reduce contributions from nonprompt leptons. 
In this analysis, "prompt leptons" refer to electrons and muons originating directly from the prompt decay of vector bosons,
while "nonprompt leptons" refer to those originating from semileptonic decays of heavy quarks, hadrons misidentified as leptons, electrons from photon conversions, 
or jets misidentified as leptons.
Charged leptons are required to originate from the PV to reject contributions from pileup   
by restricting the impact parameter (IP), defined as the distance to the point of closest approach of the track to the PV. 
The IP defined in the plane transverse to the beams is required to be $\abs{d_{xy}}<0.05\unit{cm}$, while in the longitudinal 
direction it must satisfy  $\abs{d_{z}}<0.1\unit{cm}$. 
In addition, charged leptons are required to have an IP significance (\SIP) less than 8, where 
\SIP is defined as the ratio between the IP of the associated track and its uncertainty calculated in three dimensions.
The number of electrons originating from asymmetric photon conversions is reduced by requiring  
 the track associated with each electron to have no missing hits in the innermost layers of the silicon tracker. 
An isolation variable, defined relative to the lepton \pt, is used to reduce the effect of hadronic activity in the event not connected with the lepton of interest. 
The isolation variable, \Imini,  is defined by \Imini  = $S_I(R)/\lpt$, 
with $R(\lpt)=10\GeV/\min(\max(\pt^\Pell,50\GeV), 200\GeV)$,
where $S_I(R)$ is the scalar \pt sum of charged and neutral hadrons, and photon PF candidates inside a cone with a 
variable radius $R(\lpt)$ around the lepton in the $\eta$--$\phi$ plane~\cite{Rehermann:2010vq}, and \lpt is the \pt of the lepton.
The isolation variable \Imini is calculated for each lepton, where the cone size decreases with 
increasing \lpt, reducing the probability of the overlap with a jet for Lorentz-boosted topologies.
For this analysis, muons and electrons are required to have \Imini${<}0.4$. 

A boosted decision tree (BDT) is employed for the separation of ``prompt leptons'' and ``nonprompt leptons''~\cite{CMS:2023ftu,CMS:2023dsu}. 
The BDT is trained separately for electrons and muons based on the identification methods developed for various CMS 
measurements~\cite{CMS:2020mpn,CMS:2018fdh,CMS:2018sgc,CMS:2021pqj,CMS:2021ugv} targeting the leptons from the top quark
decay chain and vector boson decays, or from particles beyond the SM~\cite{CMS:2021cox}.
Based on thresholds on the BDT discriminant value, ``loose'' and ``tight'' leptons (muons and electrons) are defined, with efficiencies of approximately 96 and 89\%, respectively.
Selected leptons are those that pass the kinematic selection criteria of $\pt>30\,(35)\GeV$ and $\abs{\eta}<2.4\,(2.5)$ for muons (electrons). 
Events are corrected for the differences between data and simulation for trigger efficiencies, lepton identification, and lepton isolation. 
The corrections are determined using the ``tag-and-probe'' method applied to \PZ boson candidate events~\cite{CMS:2011aa}
and are parameterized as functions of the lepton \pt and $\eta$.
\par}

The jets are required to have $\pt>30\GeV$ and to be within $\abs{\eta}<2.4$ for 2016 data and $\abs{\eta}<2.5$ for 2017--2018 data.
In addition, jets are required to be separated from the identified leptons by $\Delta R = \sqrt{\smash[b]{({\Delta}\eta)^2+({\Delta}\phi)^2}}>0.4$.
The charged-hadron subtraction algorithm is used to mitigate the effect of pileup~\cite{CMS:2016lmd}.
Jets with $\pt<50\GeV$ that originate from pileup events are suppressed by a ``pileup jet identification'' algorithm~\cite{CMS:2020ebo}.
For jets with $\pt>50\GeV$ no pileup identification is applied. 
Scale factors (SF) have been applied to the simulation to correct for the differences with respect to the data depending on the jet \pt and $\eta$.
The working point that corresponds to a 99\% efficiency for prompt jets and 1\% misidentification rate for pileup jets within $\abs{\eta}<2.5$ is used.
The jets originating from \PQb quarks (b jets) are tagged using the \DEEPJET~\cite{Bols:2020bkb,CMS-DP-2023-005} deep neural network algorithm 
utilizing global variables, as well as charged and neutral particle, and secondary vertex features in a jet to perform flavor tagging. 
The thresholds used for the  \DEEPJET discriminant for \PQb jets in this analysis correspond to a mistagging probability of 1\% for jets initiated by light quarks or gluons.
This choice of thresholds provides an efficiency of approximately 73-82\% for jets initiated by \PQb quarks, depending on the data-taking period, as measured in simulated multijet and \ttbar events.
To correct for the differences with respect to the data, SFs have been applied to the simulation depending on the jet flavor, \pt, and $\abs{\eta}$~\cite{CMS:2017wtu}.

The signal signature is one lepton, \ptmiss, and \PQb-tagged jets.
The main event selection requires exactly one ``tight'' lepton, either an electron or a muon, and $\ptmiss > 35\GeV$. 
Leptonic $\tau$ decays are accepted in the event selection, while hadronic $\tau$ decays are vetoed.
An additional selection on the scalar \pt sum of the jets, $\HT>200\GeV$, is applied to enhance the signal purity.
Events with additional ``loose'' leptons with $\pt>10\GeV$ are rejected.  
We define signal regions (SRs) depending on the number of \PQb-tagged jets and jets, namely 2b2j, 2b3j, 2b4j, 3b3j, and 3b4j. 
The numbers preceding b and j indicate the number of \PQb-tagged jets and jets, respectively. Regions labeled with 4j include at least four jets, while those labeled with 3b include at least three \PQb-tagged jets.
Different SRs are sensitive to different signature and mass assumptions. 
The 2b and 3b SRs are defined mainly to separate $s$-channel production from the one in association with a quark, which is predominantly a \PQb quark.
The signal selection efficiencies range from ${\approx}3$ to 30\%, depending on the SR and the assumed $\mHpm$,
with the highest efficiencies corresponding to the 2b2j SR, and the lowest efficiencies corresponding to the 3b SRs.  

\begin{figure*}[ht!]
\centering
\includegraphics[width=\textwidth]{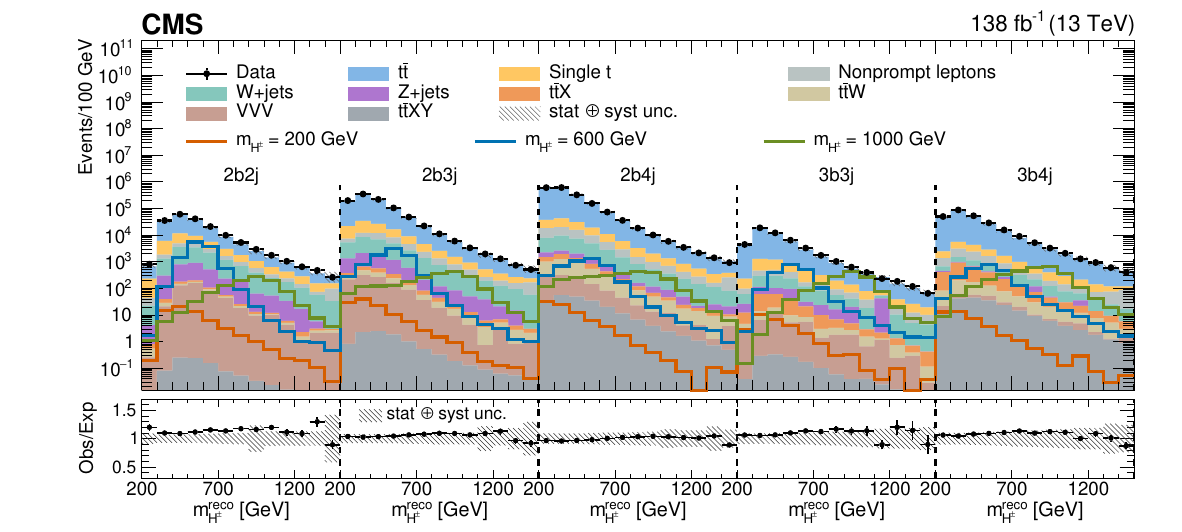}
\caption{The prefit reconstructed \Hpm boson mass distributions in each SR in the combined data set.  
Predictions for the signal with $\mHpm=200$, 600, and 1000\GeV with $\rhotc = 0.4$ and $\rhott = 0.6$, normalized to a  cross section of 1\pb, are also shown.
Beneath each plot, the ratio of data to predictions is shown.
The error bars only consider statistical uncertainties in data, while the hashed bands include both statistical and systematic uncertainties.}
\label{fig:mH_prefit}
\end{figure*}

\begin{figure*}[ht!]
\centering
\includegraphics[width=1.0\textwidth]{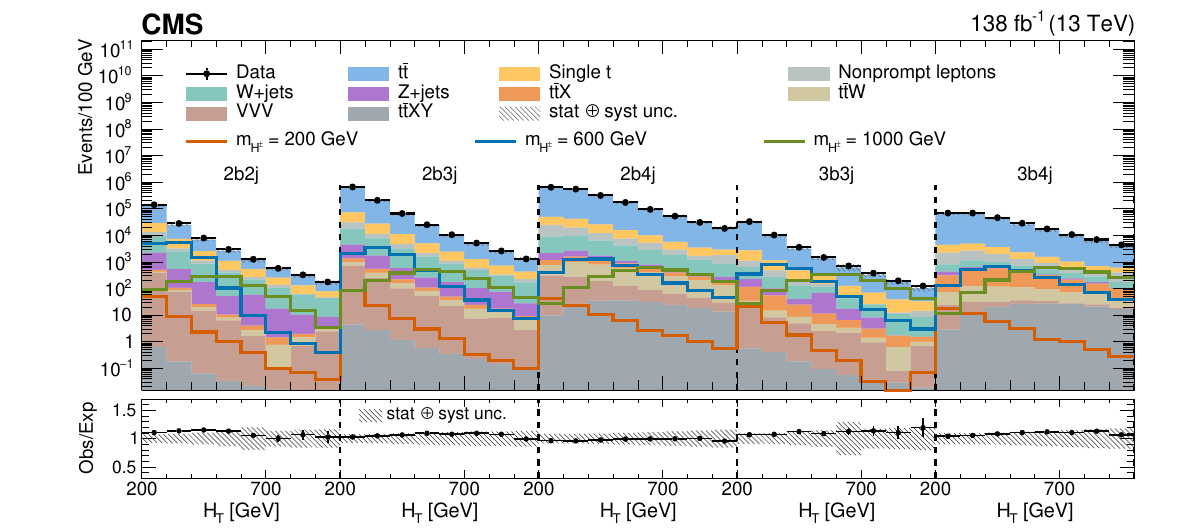}
\caption{The prefit reconstructed \HT distributions in each SR in the combined data set. 
Predictions for the signal with $\mHpm=200$, 600, and 1000\GeV with $\rhotc = 0.4$ and $\rhott = 0.6$, normalized to a  cross section of 1\pb, are also shown.
Beneath each plot, the ratio of data to predictions is shown.
The error bars only consider statistical uncertainties in data, while the hashed bands include both statistical and systematic uncertainties.}
\label{fig:HT_prefit}
\end{figure*}

For each event, the top quark and \PHpm boson masses are reconstructed.
As the signal signature contains a \PW boson originating from a top quark decay, 
the first step is to form the \PW boson from its decay products. 
The \PW boson four-momentum is obtained by summing the neutrino and lepton four-momenta. 
To achieve this, neutrino kinematic distributions are inferred using techniques described in Refs.~\cite{CMS:2021jnp,CMS:2021hug}.
It is assumed that the $x$ and $y$ components of the \ptvecmiss are entirely due to the escaping neutrino, 
and the \PW boson mass (\mW) is taken to be equal to its world average value of 80.4\GeV~\cite{ParticleDataGroup:2024cfk} to 
extract the unknown \ensuremath{z} component of the neutrino (\pznu).
In ${\approx}75\%$ of the cases \pznu has two real solutions, and we select the one that has the smaller magnitude. 
In ${\approx}25\%$ of the cases, the reconstructed transverse mass of the \PW boson is larger than \mW, 
leading to complex solutions for \pznu. 
In those cases, only the real part of the solution is kept. 
Once the neutrino and the \PW boson are reconstructed, in a second step, we evaluate the four-momentum of the top quark candidate. 
When coupled with the reconstructed \PW boson, the \PQb jet that results in a reconstructed \mt
closest to the nominal value~\cite{ParticleDataGroup:2024cfk} is chosen to be the \PQb jet candidate from the top quark decay. 
The efficiency of the top quark kinematic reconstruction, based on generator-level information about the \PQb quark origin and defined as
the fraction of events for which the reconstructed \PQb jet is correctly assigned to the top quark it originates from, 
varies between 61 and 94\% for the 2b SR, and between 53 and 81\% for the 3b SR, depending on \mHpm.
The \PQb jet candidate combined with the top quark candidate that gives the highest \mHpm is selected to be the \PQb jet 
from the \Hpm boson decay. The reconstruction efficiency of \PHpm rises from 27\% at $\mHpm=200\GeV$ to 93\% at $\mHpm=1\TeV$.
The reconstructed distributions of the two discriminating variables, \mHpm and \HT, 
in the 2b2j, 2b3j, 2b4j, 3b3j, and 3b4j regions are displayed in Figs.~\ref{fig:mH_prefit}--\ref{fig:HT_prefit}. 
A good agreement is observed between data and the sum of the background predictions within the uncertainties. 
Backgrounds are detailed in Section~\ref{sec:bkgd}.

\section{Background estimation from control samples in data}
\label{sec:bkgd}
Several processes can mimic the signature of the signal events. After selection, these backgrounds are dominated by \ttbar and single top quark, and \wjet production,
and by jets misreconstructed as leptons (``nonprompt''). 
Minor contributions come from \zjet, \ttW, \ttX (\ttZ and \ttH) events and  
even smaller contributions from \ttXY (\ttHH, \tttj, \tttt, \tttW, \ttWH, \ttWW, \ttWZ, \ttZH, and \ttZZ events), and \VVV production are 
considered as well.
All backgrounds, except for the nonprompt-lepton contribution, are estimated using simulated event samples. 

The nonprompt-lepton background is derived from data control regions using the ``tight-to-loose'' ratio method
(discussed, e.g., in Refs.~\cite{Chatrchyan:2011wba,Chatrchyan:2012ira,Chatrchyan:2012paa,Chatrchyan:2012sa,Khachatryan:2016kod}).
The nonprompt leptons originate from heavy-flavor hadron decays, misidentified hadrons, decay products of light mesons, 
or photon conversions inside jets.
This method requires ``tight'' and ``loose'' lepton candidates that correspond to signal and nonprompt leptons, respectively.
The muon and electron misidentification rates are measured using the following selection: exactly one muon or electron that passes 
the ``tight'' object selection or ``loose'' selection, 
no additional muons or electrons that pass the ``loose'' selection, $\ptmiss<35\GeV$, and exactly one \PQb-tagged jet.
Four orthogonal regions are defined: 
A = signal region ($\ptmiss>35\GeV$, ``tight'' lepton), B = ($\ptmiss>35\GeV$, ``loose'' lepton), C = ($\ptmiss<35\GeV$, ``tight'' lepton), 
and D = ($\ptmiss<35\GeV$, ``loose'' lepton).
A multijet-enriched sample is selected from data, from which the ratio of events ($\alpha$) in region C to D is obtained in bins of \pt and $\abs{\eta}$ of the lepton,
and the scale factor $f_p$ is then calculated as $f_p = \alpha/(1-\alpha)$.
The nonprompt-lepton background contribution in the signal region (region A) is then estimated by applying $f_p$ to events in a data
control sample (region B). 
The dominant contribution of the nonprompt-lepton background originates from QCD multijet events.
A 50\% uncertainty in the nonprompt-lepton background rate is assigned.  
We validate the ``tight-to-loose'' method by performing a closure test: the nonprompt-lepton background estimated from data (region D) with this procedure is compared with simulated multijet events in region C. 
In addition, the consistency of the overall nonprompt-lepton background estimate is checked using events in an independent region with at least 4 jets, with one of them being \PQb tagged. 

\begin{table*}[!htpb]
  \centering
  \topcaption{Input variables of the \pDNNtwob and \pDNNthreeb. Indices $i < j$ run  up to including 2 for \pDNNtwob and 3 for \pDNNthreeb.}
  \begin{tabular}{ll}
  $\pt(\Pell)$   &   Transverse momentum  of the lepton\\
  \ptmiss &  Missing transverse momentum\\
  $\Delta R(\PQb_i,\Pell)$ &  Angular separation between \PQb jet and lepton\\
  $\Delta R(\PQb_i,\PQb_j)$ &  Angular separation between \PQb jets\\
  $\pt({\rm \PQb}_i)$ &  Transverse momentum  of \PQb jets\\
  \HT &  Scalar \pt sum of the jets\\
  $m(\Pell,\PQb_i,\PQb_j)$  & Invariant mass of \PQb jets and lepton\\
  $m(\Pell, \PQb_i)$ &  Invariant mass of \PQb jet and lepton\\
  $\Delta \phi(\ptmiss, \Pell)$ &  Azimuthal angle difference between missing \\
                               &  transverse momentum and lepton\\
  $\mHpm^\text{reco}$ &  Invariant mass of the reconstructed \Hpm \\
  \end{tabular}
  \label{tab:DNN_inputs}
\end{table*}

 \section{Signal extraction}
 \label{sec:signalextract}
Distributions of parameterized deep neural network (pDNN)~\cite{Baldi:2016fzo} scores are used to separate signal and background events. 
The pDNN input features used in the analysis in the 2b (\pDNNtwob) and 3b (\pDNNthreeb) SRs are listed in Table~\ref{tab:DNN_inputs}.
The input features are selected based on the correlation between features and the results of the Boruta algorithm~\cite{Kursa2010BorutaA}.
This algorithm shuffles the original features to create copies of them, called shadow features, and trains a random forest classifier on the extended data set.
The features that have lower variable importance than the highest rated shadow feature are regarded as unimportant and are removed.
Consistent features in the 2b and 3b regions are selected for simplicity, with no loss in performance.
In the pDNN, the neural network is parameterized by \mHpm, which is taken as an additional input variable of the network.
This feature is assigned to be the generated \mHpm for the signal and assigned randomly from the \mHpm values for the background.
The pDNN interpolates between different \mHpm hypotheses. 
Additionally, it provides effective interpolation for the mass input to pDNN for modeling the background. 
However, for the signal, since the input kinematic distributions change in a nontrivial way, interpolation can not fully capture the differences. 
Therefore, we provide results on specific mass hypotheses. 
Separate trainings are performed for 2016 and 2017--2018 data to take into account the changing detector conditions. 
The pDNNs are trained separately for the 2b2j, 2b3j+2b4j, 3b3j, and 3b4j SRs and the two data-taking periods.  
The signal is extracted from the pDNN score distributions combining all data-taking periods in each SR.  
All signal and background processes are considered for the pDNN training, except for the nonprompt-lepton background.
The importance of the input features varies across the SRs.
In general, the most sensitive variables are related to the leading \PQb jet \pt, 
the invariant masses of the lepton and the \PQb jets, and \HT.

Reweighting is performed among signal samples, so that each mass hypothesis enters with the same weight.
The weight assigned to each background sample during the training is proportional to its expected event yield.
The pDNN design contains three hidden layers.
Hyperparameter tuning is performed using part of the training data set to minimize the loss function, which is chosen to be the binary cross entropy.
The \ensuremath{k}-fold (\ensuremath{k} = 5) strategy  is used to avoid overtraining. 
We first split the data set into an application sample and a training+validation sample.
Then, the training+validation sample is used for \ensuremath{k}-fold training, so the application sample remains
independent of the training and validation. 
Using the Adam algorithm~\cite{kingma2017adammethodstochasticoptimization} employing the Nesterov momentum~\cite{dozat.2016} to minimize the binary cross entropy 
loss function, the model is trained for 250 epochs.
The final model is the ensemble of the models produced from \ensuremath{k}-fold training weighted with inverse of its loss in the \ensuremath{k}-fold test.
Receiver operating characteristic curves in each fold 
and the comparison between validation and training distributions are used to check the performance and potential overtraining of the pDNN.

The bins of the pDNN score distribution are optimized in several steps. 
First, the pDNN distribution is constructed with 10 bins of equal size. 
Second, the bins are combined so that the yield in each bin is at least 10 events. 
In the next step, the pDNN distribution is rebinned using the Bayesian block algorithm~\cite{BBA_code,2013ApJ...764..167S} based on the signal distribution,
to ensure the stabi\-lity of the signal extraction procedure. 
Further optimization is performed in order to avoid large differences of bin widths between two consecutive bins. 
This last step of the optimization procedure is not applied to the lowest and highest bins.  

The low pDNN score signal region (pDNN${<}0.5$) is used to further validate the ``tight-to-loose'' method 
by comparing the observations with the full background prediction, including the estimated nonprompt-lepton contribution.

\section{Systematic uncertainties}
\label{sec:systematics}

Systematic uncertainties, arising from various sources, such as detector effects, theoretical uncertainties, and mismodeling,
are evaluated and categorized in two main groups: experimental and modeling uncertainties.   
The systematic uncertainties are modeled as nuisance parameters in the maximum likelihood estimation procedure 
used for determining the best fit signal strength $\hat{\mu}$, defined as the ratio of the measured cross section to the signal expectation in the g2HDM model.
The nuisance parameters can modify the shape and normalization of the background and signal distributions in the fit using the CMS statistical analysis tool \textsc{Combine}~\cite{CMS:2024onh}.
The chosen probability density function (pdf) used for the interpolation between nominal and modified templates is dependent on whether the associated nuisance 
parameter can also affect the shape of the nominal template or not. Log-normal pdfs are used for nuisance parameters that only affect the distribution normalizations, 
whereas Gaussian pdfs are used for those that can affect the shape as well.
Experimental uncertainties, where appropriate, and all modeling uncertainties are taken as correlated across the data-taking periods, SRs, and channels. 
The effect of the limited number of simulated events is estimated using the Barlow--Beeston ``light'' method~\cite{Barlow:1993dm}. 

The experimental uncertainties are described below.
The integrated luminosity measurements of CMS for the 2016, 2017, and
2018 data-taking periods are described in
Refs.~\cite{CMS:2021xjt,CMS-PAS-LUM-17-004,CMS-PAS-LUM-18-002}. The
uncertainty in the integrated luminosity has an effect on the
background and signal yields. The combined uncertainty for 2016--2018 is
1.6\%. To estimate the effect of mismodeling of the pileup, the
total inelastic cross section of 69.2 mb is varied by
$\pm4.6\%$~\cite{CMS:2018mlc}. The variations are treated as fully correlated between the
different data-taking years. In 2016 and 2017, a time jitter of the
trigger signal in the
muon~\cite{CMS:2021yvr} and ECAL~\cite{CMS:2020cmk} detectors, led to
a loss of events of the order of ${\lesssim}0.5\%$. 
These effects are corrected for by applying dedicated SFs to the simulation. 
The associated uncertainties are evaluated by varying the SFs within their uncertainties.

The trigger SF uncertainties are assumed to be uncorrelated
between the data-taking periods. The uncertainties in the electron
reconstruction efficiency SFs are considered to be
correlated across the data-taking years. For the uncertainties in the SFs
related to lepton identification, isolation, energy scale, and
resolution, the statistical and systematic components are merged, and
therefore, for both muons~\cite{CMS:2018rym} and electrons~\cite{CMS:2020uim}, nuisance parameters are considered
correlated across years.
  
{\tolerance=1200
For the \PQb tagging and misidentification efficiencies, two nuisance parameters are
considered for light-quark/gluon and heavy-flavor jets~\cite{CMS:2017wtu}. 
Each of them have components correlated/uncorrelated across the data-taking periods. 
The variations of jet four-momenta using the full set of
uncertainty sources and corresponding correlations from
Ref.~\cite{CMS:2016lmd} are used. The uncertainties in the jet energy resolution in
different data-taking periods are considered uncorrelated, and in different
channels fully correlated. The uncertainties in jet energy scale and resolution, as well
as the variation of the unclustered energy, are propagated into
\ptmiss. The uncertainty in \ptmiss is taken as uncorrelated across the
three data-taking years. 
For the jet pileup identification scale factors, one nuisance parameter is calculated 
from three components combined in quadrature: the statistical uncertainty, the uncertainty 
in the efficiency extracted from simulated events, and the jet energy scale uncertainty. 
The uncertainty in the jet pileup identification is taken as correlated across channels, the 
data-taking years, and the SRs. 
\par} 
  
Finally, two separate nuisance parameters are used for the nonprompt-lepton
background estimation from data: a normalization uncertainty of 50\% and a shape
uncertainty estimated from the difference between the 1\PQb
and 2\PQb nonprompt-lepton background distributions.  The normalization uncertainty is
uncorrelated across the SRs and channels.
  
In the following, we describe the sources of modeling uncertainty. 
The uncertainties in the signal, \ttbar, and \wjet backgrounds are included as
separate nuisance parameters that affect the shape of the distributions, and additional
parameters are included that affect their normalization.  
For the other backgrounds, which are estimated using
simulated event samples, the corresponding uncertainties in the
cross section calculations are used, and nuisance parameters are included that only
affect the yields.
As discussed in Section~\ref{sec:sim}, the simulated \ttbar events are reweighted to account for the difference between the top quark \pt distribution in data and simulation.
For the reweighting, which only affects the \ttbar background, uncertainties related to the correction from NLO 
to NNLO predictions are estimated using the difference between the
NNLO and NLO predictions. These uncertainties are treated as uncorrelated
across the SRs.

Variations of \mur and \muf by factors of 0.5 and 2 are considered and
assumed uncorrelated with each other. 
This uncertainty is taken correlated across years and channels. The uncertainties due
to the PDF set used in the simulation are estimated using the Hessian
variations of the NNPDF3.1 NNLO set. The effects are propagated through
event weights~\cite{Butterworth:2015oua}. The effect of the variation
of \alpS on the PDF set is also taken into account. The
PDF uncertainties are considered fully correlated across the
data-taking years. The parton shower scale choice is varied by a
factor of 0.5 and 2 for initial-state and final-state radiation separately, using event weights for the signal,~\ttbar, and
\wjet samples.  This uncertainty is
considered correlated for the same process across the years and
channels.

 \section{Results}
\label{sec:results}

\begin{figure*}[ht!]
\centering
\includegraphics[width=0.4\textwidth]{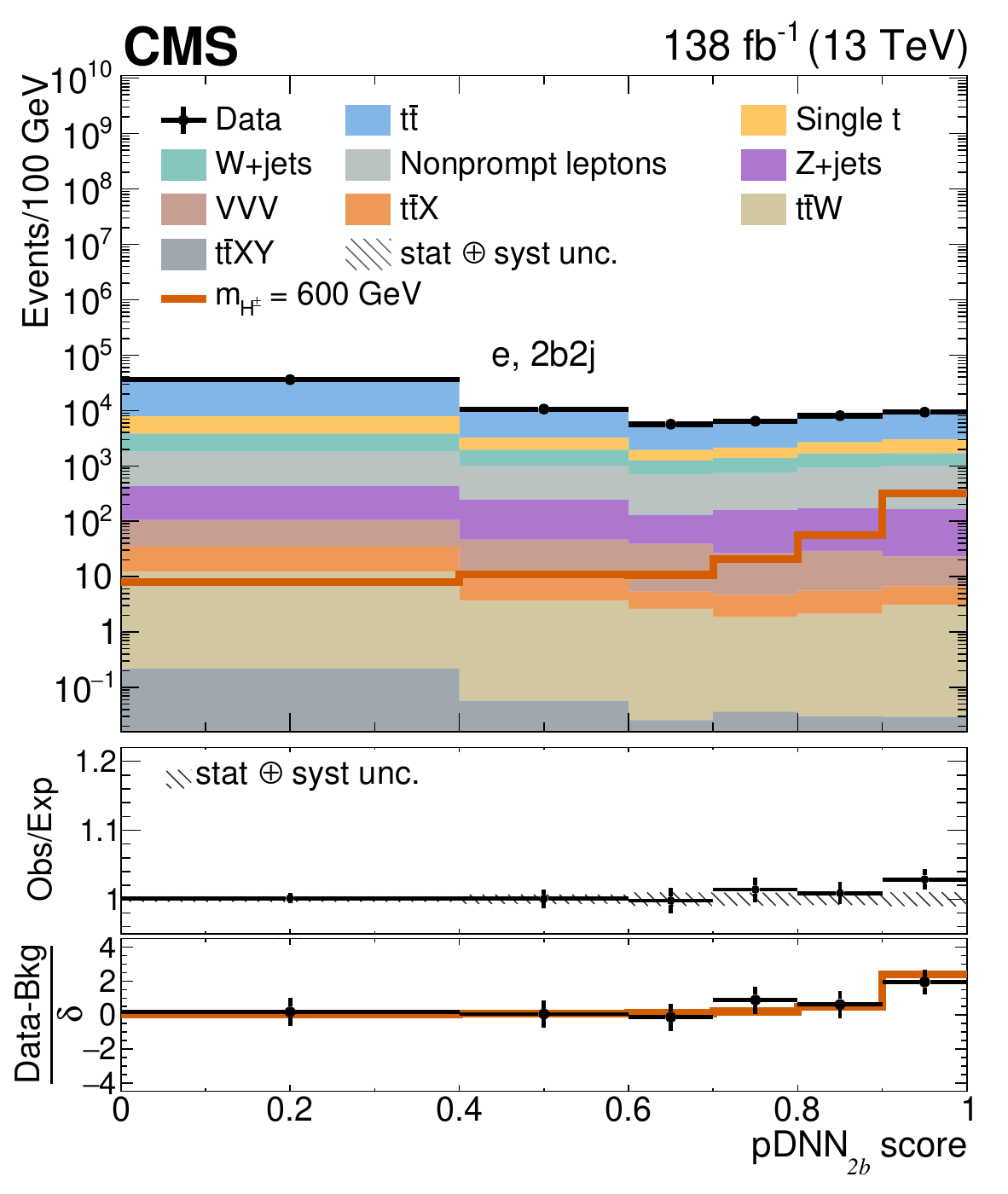}
\includegraphics[width=0.4\textwidth]{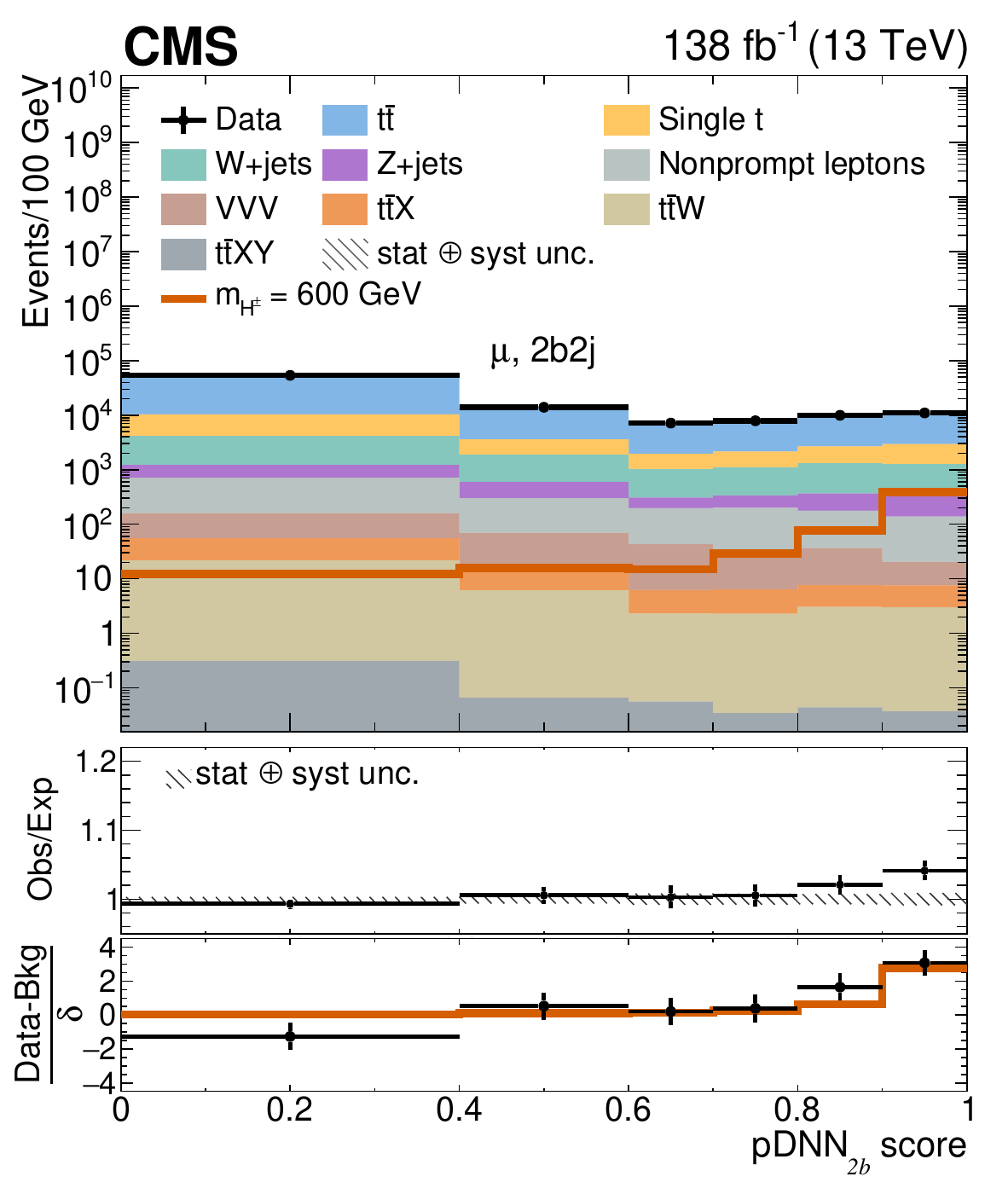}
\includegraphics[width=0.4\textwidth]{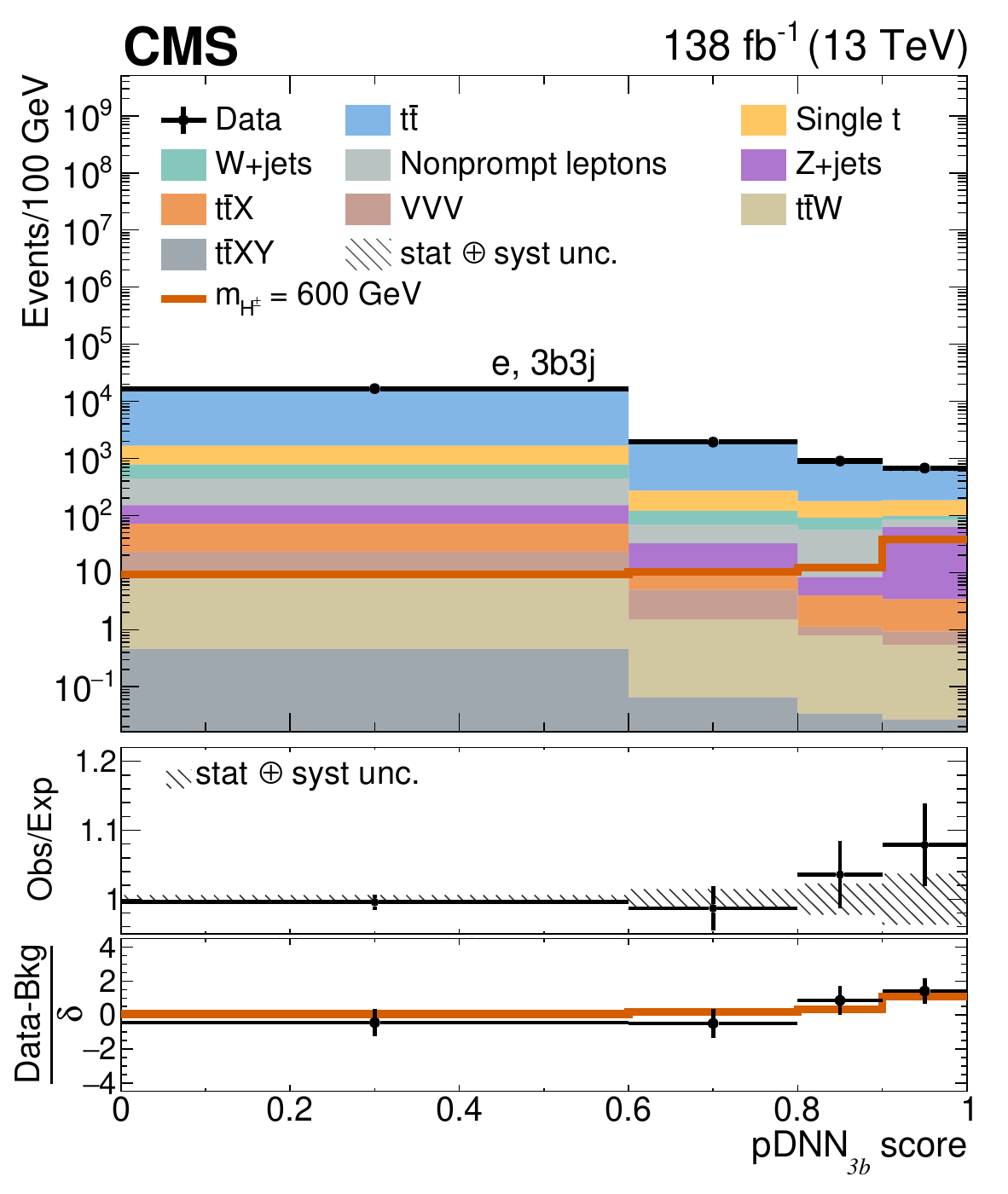}
\includegraphics[width=0.4\textwidth]{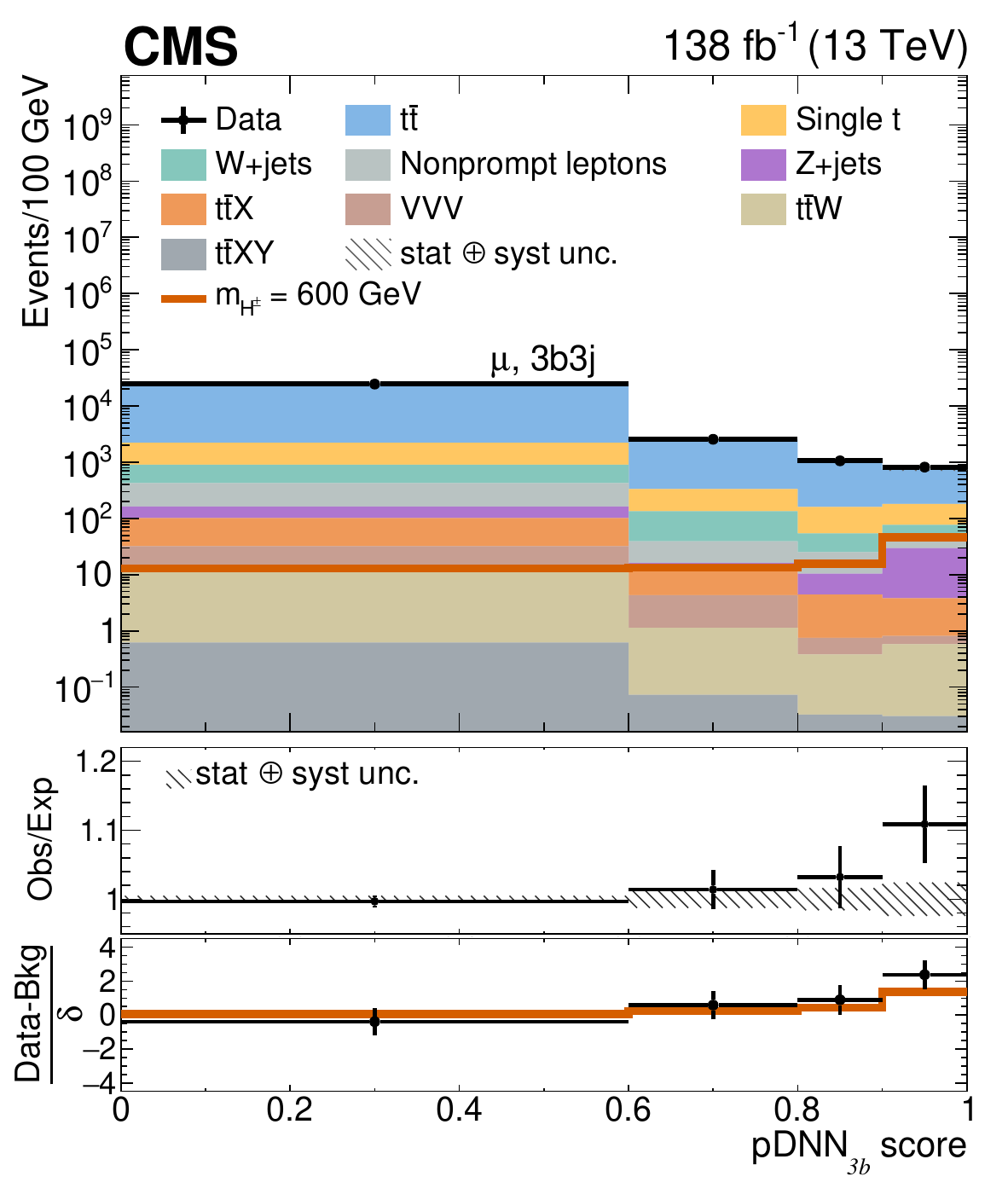}
\caption{The postfit pDNN distributions in the SR 2b2j (upper) and 3b3j (lower) for electron (left) and muon (right) channels assuming $\mHpm=600\GeV$. 
Postfit signal for $\mHpm=600\GeV$ with $\rhotc = 0.4$ and $\rhott = 0.6$ is also shown.
Lower panels display the difference between data and the sum of the SM background predictions, normalized by the combined statistical and systematic uncertainty, $\delta$.
The error bars only consider statistical uncertainties in data, while the hashed bands include both statistical and systematic uncertainties.}
\label{fig:postFit}
\end{figure*}

\begin{figure*}[ht!]
\centering
\includegraphics[width=0.4\textwidth]{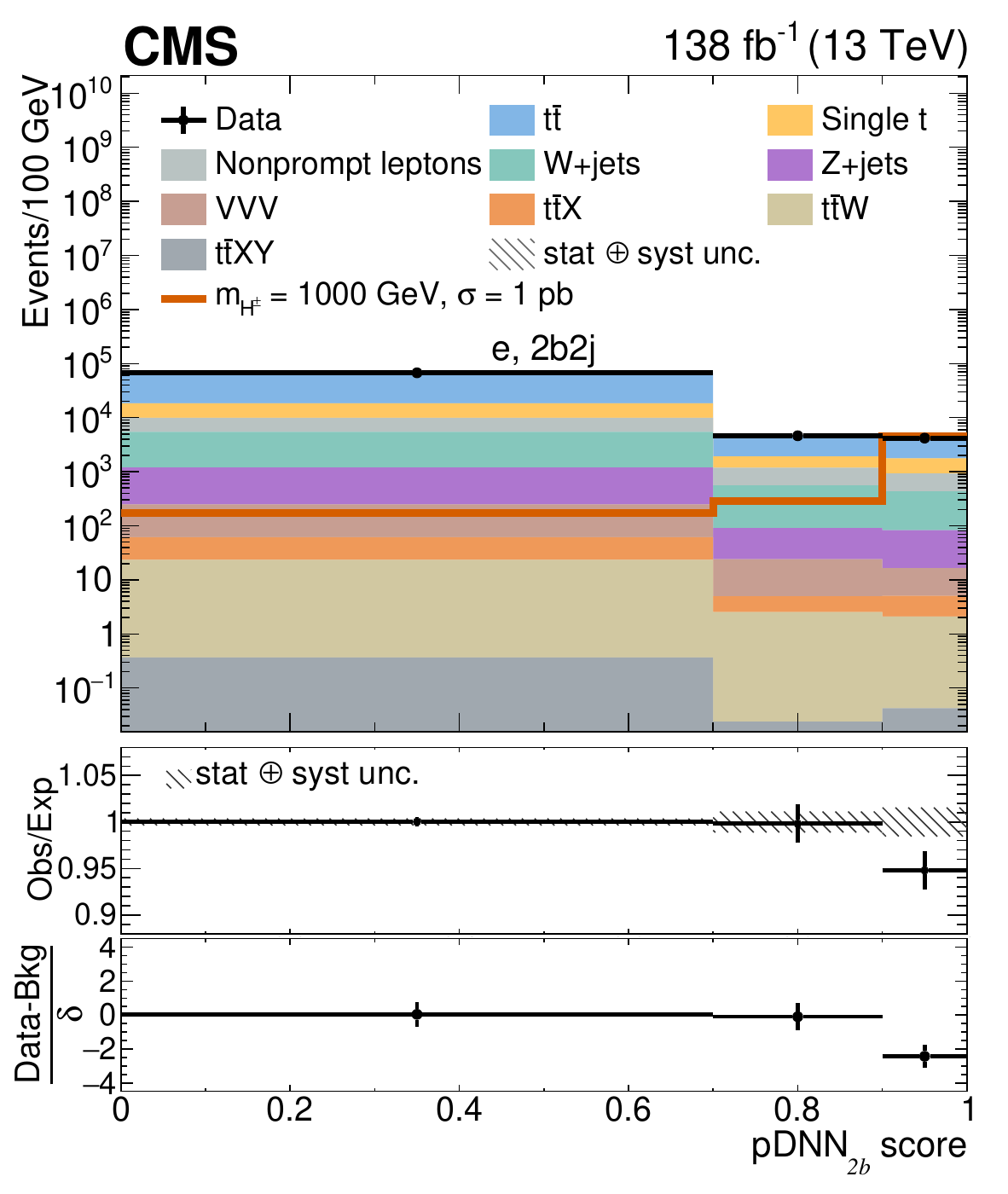}
\includegraphics[width=0.4\textwidth]{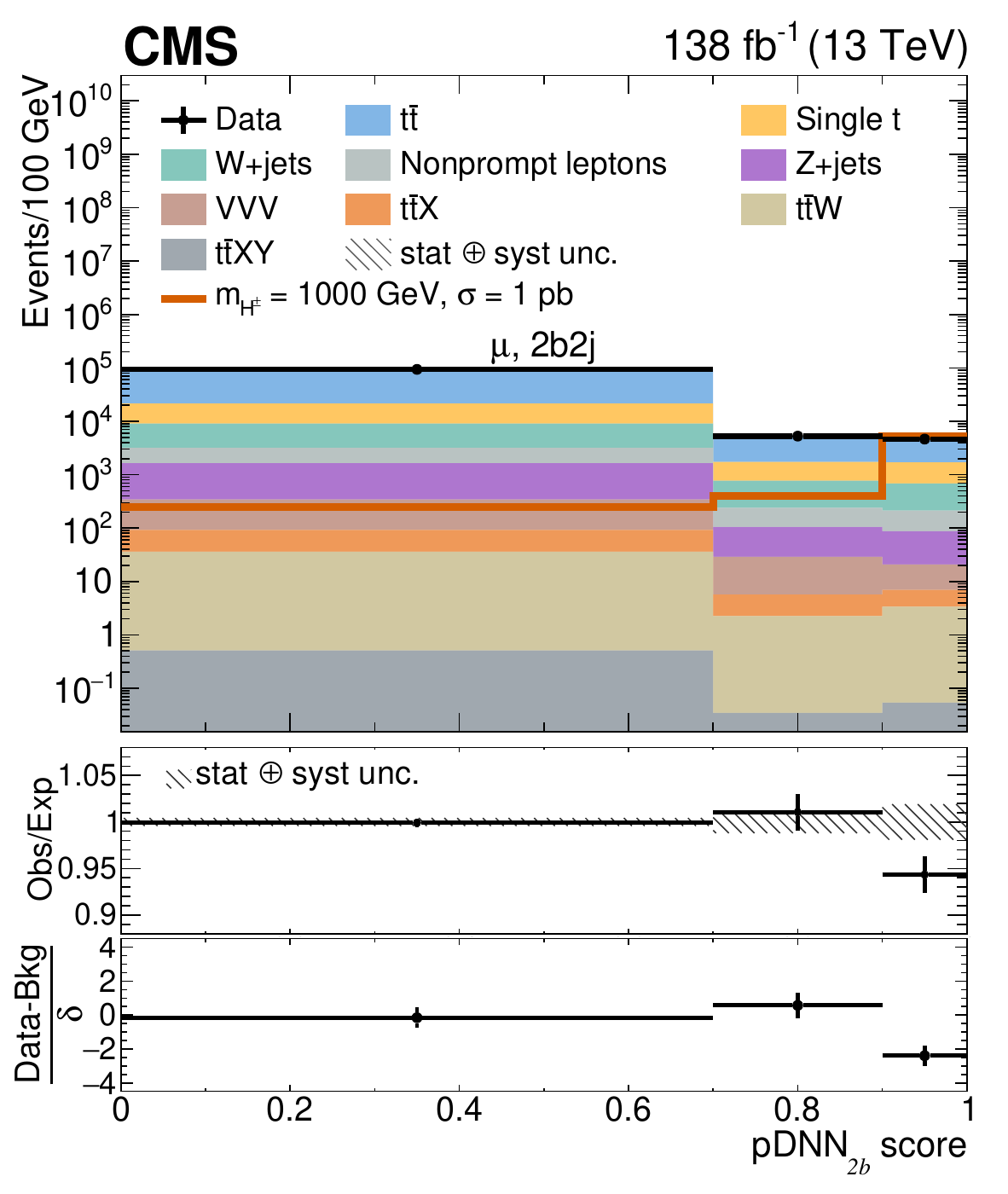}
\includegraphics[width=0.4\textwidth]{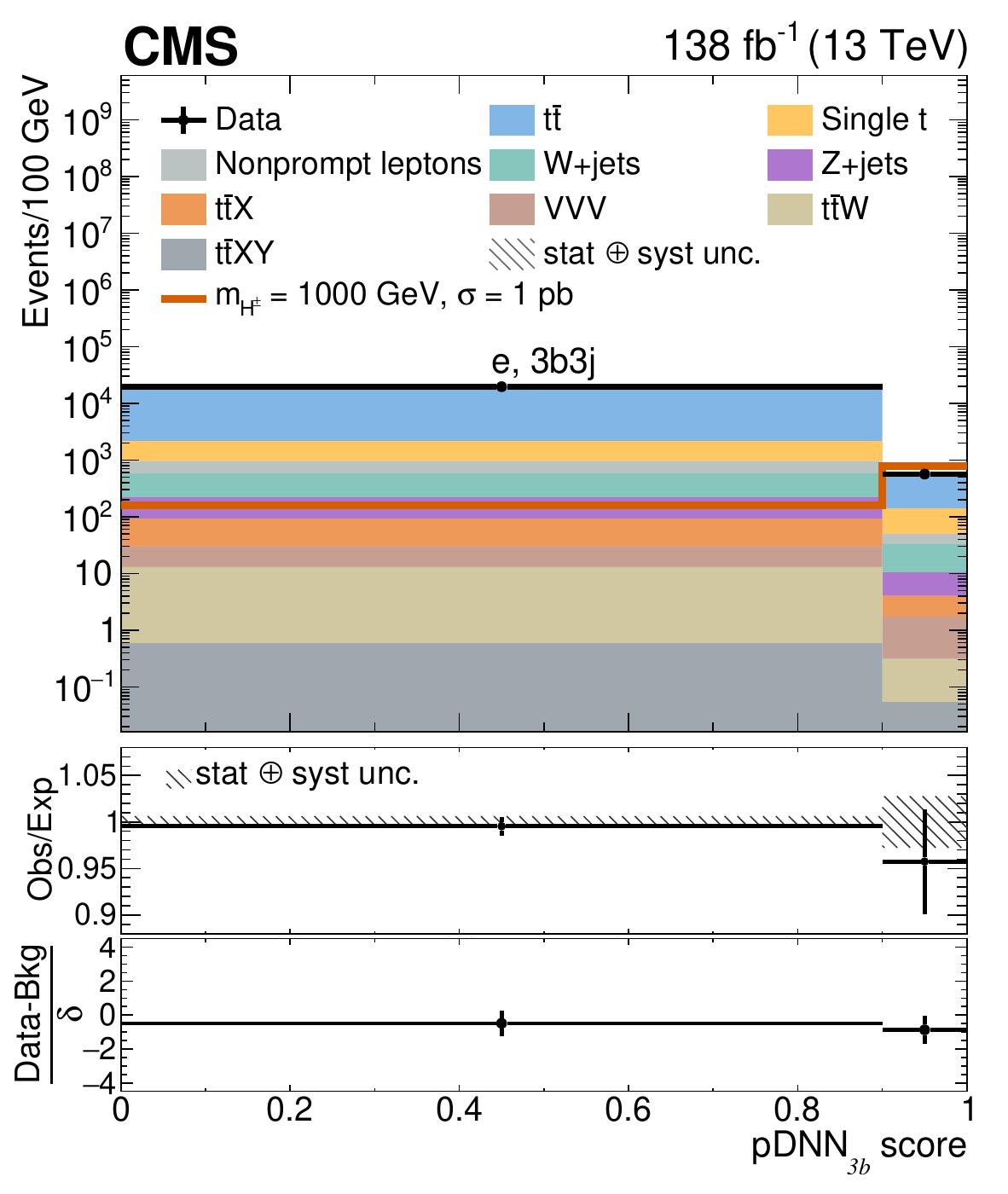}
\includegraphics[width=0.4\textwidth]{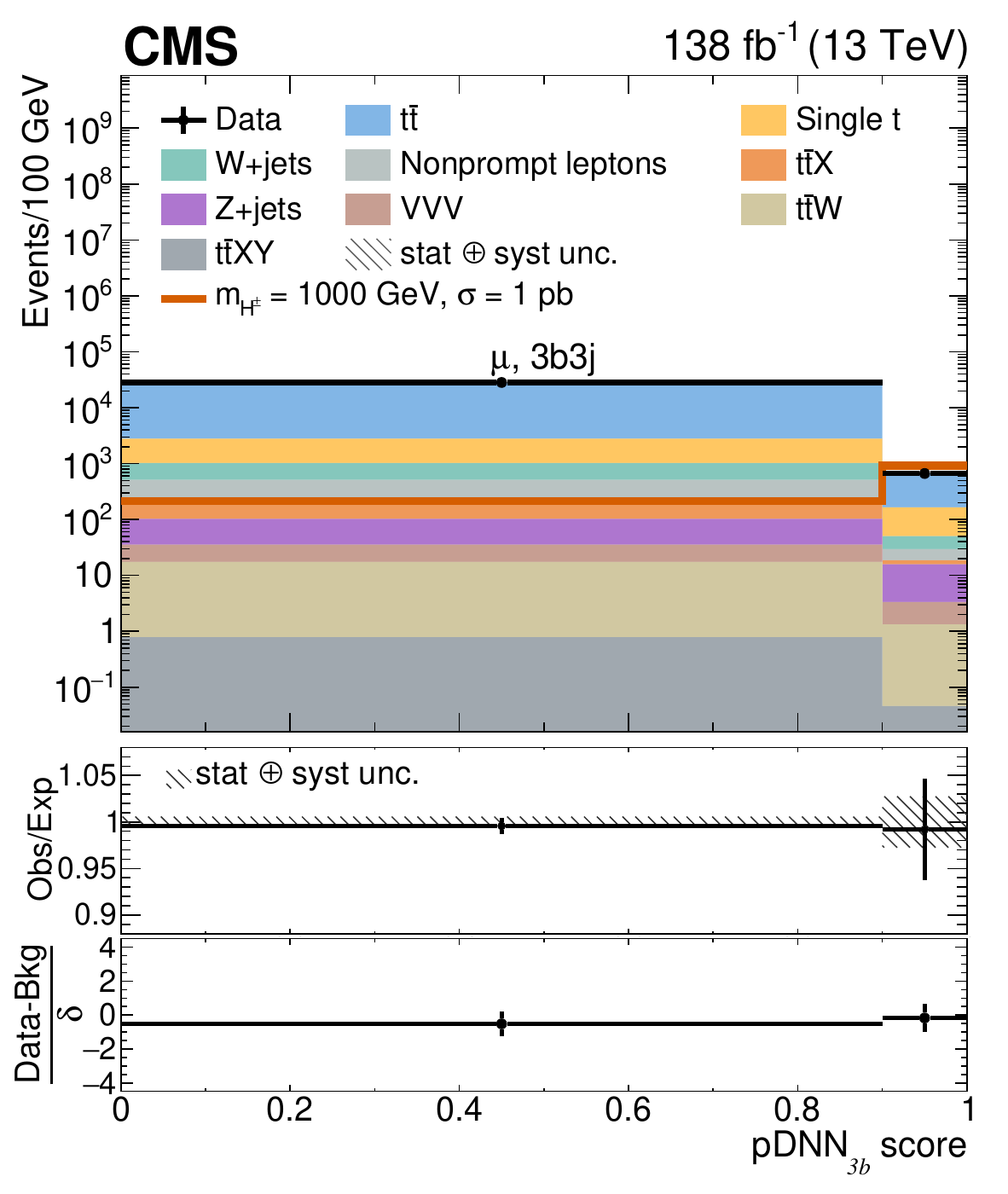}
\caption{The postfit pDNN distributions in the SR 2b2j (upper) and 3b3j (lower) for electron (left) and muon (right) channels assuming $\mHpm=1\TeV$. 
The signal for $\mHpm=1\TeV$ with $\rhotc = 0.4$ and $\rhott = 0.6$ is also shown before the fit, assuming a cross section of 1\pb.
Lower panels display the difference between data and the sum of the SM background predictions, normalized by the combined statistical and systematic uncertainty, $\delta$.
The error bars only consider statistical uncertainties in data, while the hashed bands include both statistical and systematic uncertainties.}
\label{fig:postFit1000}
\end{figure*}

The signal is extracted by simultaneously fitting the pDNN distributions across two channels (\Pe and \PGm), 
and five SRs (2b2j, 2b3j, 2b4j, 3b3j, and 3b4j) combining all data-taking periods. 
For the background processes, the yields and shapes are taken directly from the simulation except for 
the nonprompt-lepton background. 
Despite having a small signal contribution, the 2b4j region is included to constrain the background-related nuisance parameters from data directly in the fit. 
Figures~\ref{fig:postFit} and \ref{fig:postFit1000} display the postfit pDNN discriminant distributions 
used for the signal extraction in the 2b2j and 3b3j SRs for the \Pe and \PGm channels, 
for $\mHpm=600$ and 1000\GeV, respectively. 
In Fig.~\ref{fig:postFit}, the postfit distribution for the signal at  $\mHpm=600\GeV$, which displays the highest deviation with respect to the SM, is shown. 
The postfit signal distribution for $\mHpm=1\TeV$ displays zero signal yield as a result of the fit; hence in Fig.~\ref{fig:postFit1000} 
the  distribution is shown for the signal at $\mHpm=1\TeV$ before the fit, assuming a cross section of 1\pb.

A summary of the decomposition of the systematic uncertainties with respect to the total uncertainty in the signal strength for $\mHpm=200, 600$, 
and 1000\GeV is shown in Tables~\ref{tab:syst_rank_H200}--\ref{tab:syst_rank_H1000}. 
Dominant systematic uncertainties affecting the results depend on the signal mass. 
In general, theoretical background cross sections, top quark \pt reweighting, heavy-flavor tagging, and statistical uncertainties 
have the largest contributions to the total uncertainty in the signal strength.  
At higher $\mHpm$ values, the signal and background discrimination becomes more efficient, therefore the relative impact of the theoretical MC uncertainties in the fit is reduced. 
While both heavy-flavor tagging and PDF uncertainties rise from 200 to 600 GeV, the relative impact of the heavy-flavor tagging nuisance plateaus thereafter,
but the PDF uncertainty continues to grow as partonic momentum fractions probe less constrained phase space regions.

\begin{figure*}[ht!]
  \centering
  \includegraphics[width=0.51\textwidth]{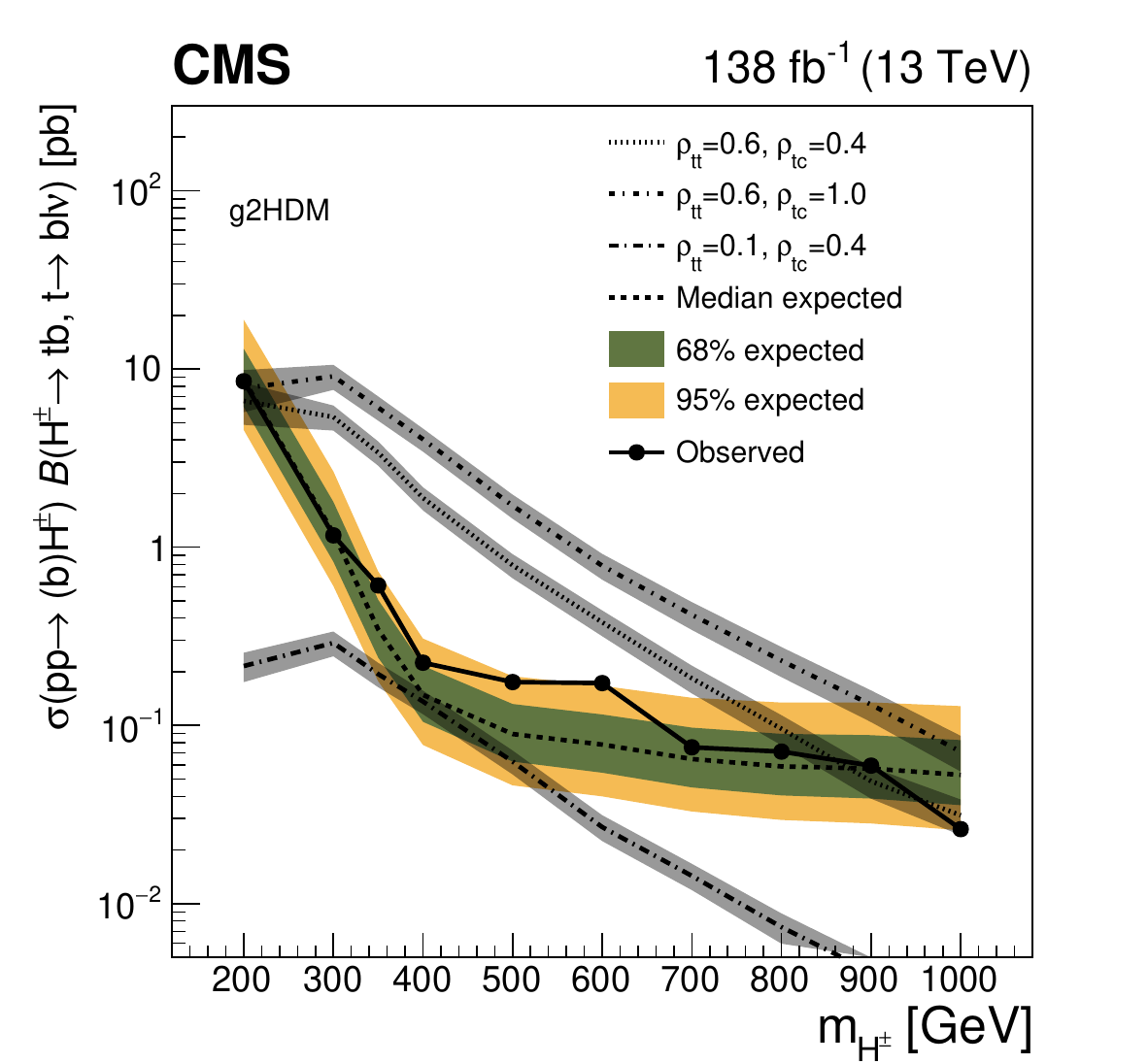}
  \includegraphics[width=0.48\textwidth]{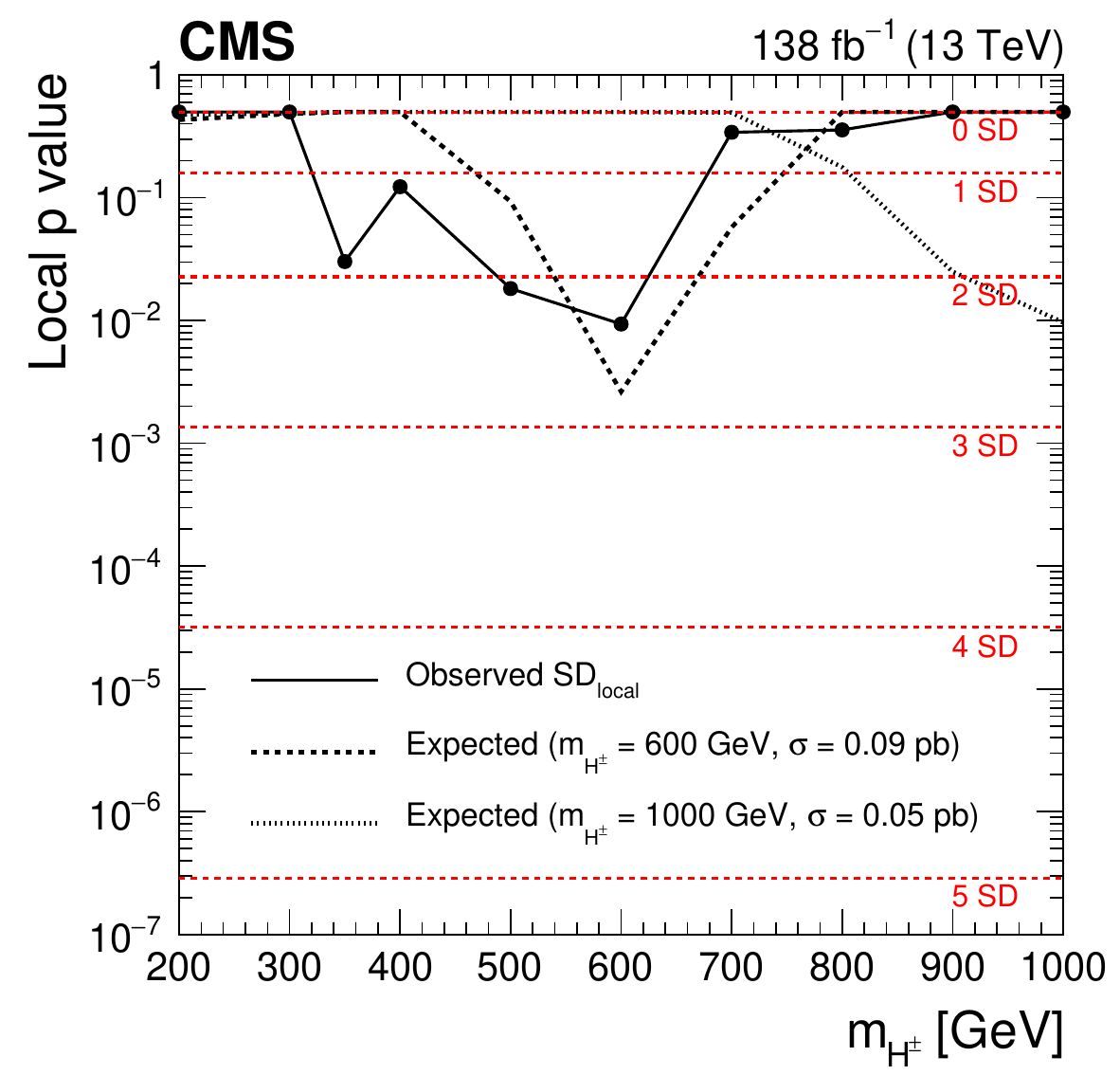}
  \caption{Left: observed and expected 95\% \CL limits on the cross section times branching fraction ($\sigma\mathcal{B}$) for different \Hpm boson mass hypotheses with $\rhotc = 0.4$ and $\rhott = 0.6$. 
  The inner (green) band and the outer (yellow) band represent the regions containing 68 and 95\%, respectively, of the distribution of limits expected under the background-only hypothesis.
  Theoretical prediction with different $\rhotc$ and $\rhott$ couplings are shown 
  with the grey bands representing the corresponding uncertainties due to the factorization and renormalization scales, and the PDFs.
  Right: Observed local significance extracted from the pDNN distribution. 
  Expected significances are obtained with the injection of a $\sigma=0.09\,(0.05)\pb$ of g2HDM signal with $\mHpm=600\GeV\,(1\TeV)$. 
  The injected cross sections are picked as the best fit value for $\mHpm=600\GeV$, and the theoretical prediction for $\mHpm=1\TeV$ assuming $\rhotc=\rhott=0.6$.
  Local significance values in units of standard deviations (SD) are also shown with dashed horizontal lines on the  vertical axis.
  }
  \label{fig:limit_run2_DNN}
\end{figure*}

\begin{figure*}[htb!]
  \centering
  \includegraphics[width=0.45\textwidth]{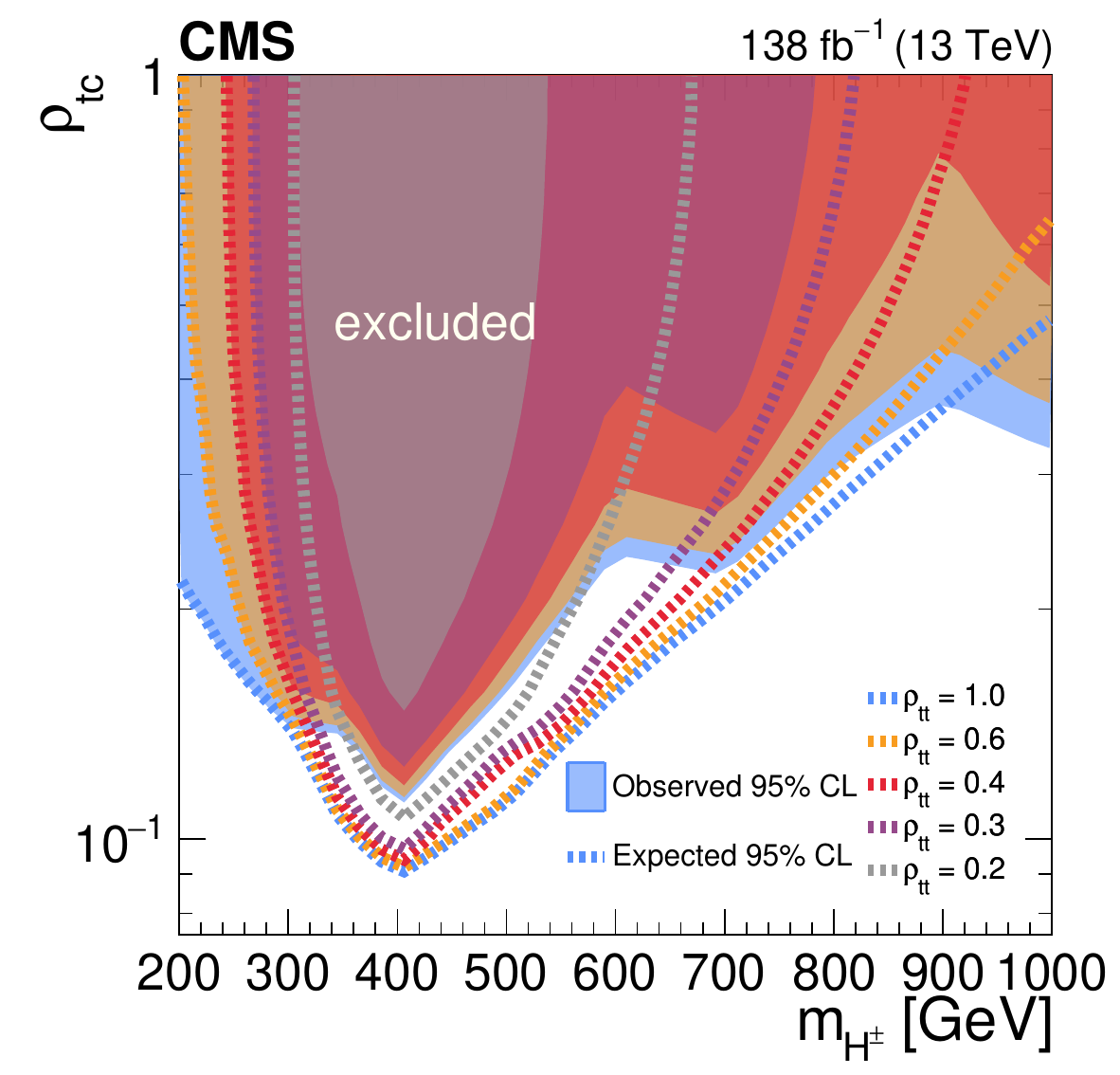}
  \caption{Observed and expected 95\% \CL excluded phase space regions, as functions of \mHpm and \rhotc for various assumed \rhott values represented with different colors.  
  The limits are extracted from the pDNN distributions based on the g2HDM assuming 
  all the extra Yukawa couplings except $\rhott$ and $\rhotc$ are zero. 
  The results are obtained from the 95\% \CL limits on the cross section times branching fraction in Fig.~\ref{fig:limit_run2_DNN}.}
  \label{fig:exclusion_run2}
\end{figure*}

The results of the search are interpreted in terms of upper limits on the product of the cross section 
\crosssectionbHwPar and branching fraction $\mathcal{B}(\htb, {\PQt\to\PQb \Pell\PGn})$ in the g2HDM, where $\Pell=\Pe, \PGm$.
The upper limits on cross sections are calculated at 95\% confidence level (\CL) using a modified frequentist 
method~\cite{Heinemeyer:2013tqa, bib:CLS2, bib:CLS1} within an asymptotic approximation~\cite{Cowan:2010js}.
The expected 95\% \CL on the production cross section and the regions containing 68 and 95\% of the distribution of the 
limits expected under the background-only hypothesis are calculated with pseudo-experiments using background-only samples. 
Modifying the extra Yukawa couplings in the range 0.1--1.0 does not impact the limits significantly, and they remain within the 68\% \CL band. 
The \Hpm boson width increases with the mass and the strength of the extra Yukawa couplings. 
The analysis is not sensitive to the width up to $\sim$10\%, and the largest \Hpm boson width in the signal phase space 
probed in this analysis is ${\sim}12\%$, which is obtained for $\mHpm=1\TeV$ and $\rhott=\rhotc=1$. 
Figure \ref{fig:limit_run2_DNN} displays the observed and expected 95\% \CL upper limits on the cross section times branching fraction, 
along with the corresponding observed and expected local significances, $p$-value~\cite{Cowan:2010js}, as a function of \mHpm. 
The smallest $p$-value is observed for a signal with $\mHpm=600\GeV$, corresponding to a local significance of 
2.4 standard deviations (SD). 
Figure~\ref{fig:exclusion_run2} displays the excluded phase space regions as functions of \mHpm and \rhotc for various assumed \rhott values   
based on the g2HDM with all the extra Yukawa couplings except $\rhott$ and $\rhotc$ set to zero. 
To obtain this distribution, a linear interpolation is applied to the exclusion limits over 20 \rhotc coupling points in the range 0.05 to 1.0, with 10 mass points included in the analysis for the final interpretation of the signal phase space.
The largest exclusion region is observed around $\mHpm=400\GeV$, excluding values of $\rhotc>0.11\,(0.12)$ for $\rhott=1.0\,(0.3)$ at 95\% \CL.
For $\rhott=1$, values of $\rhotc>0.38$ are excluded at 95\% \CL for the entire mass range tested, for $\mHpm$ from 200 to 1000\GeV.

\begin{figure*}[ht!]
\centering
\includegraphics[width=0.75\textwidth]{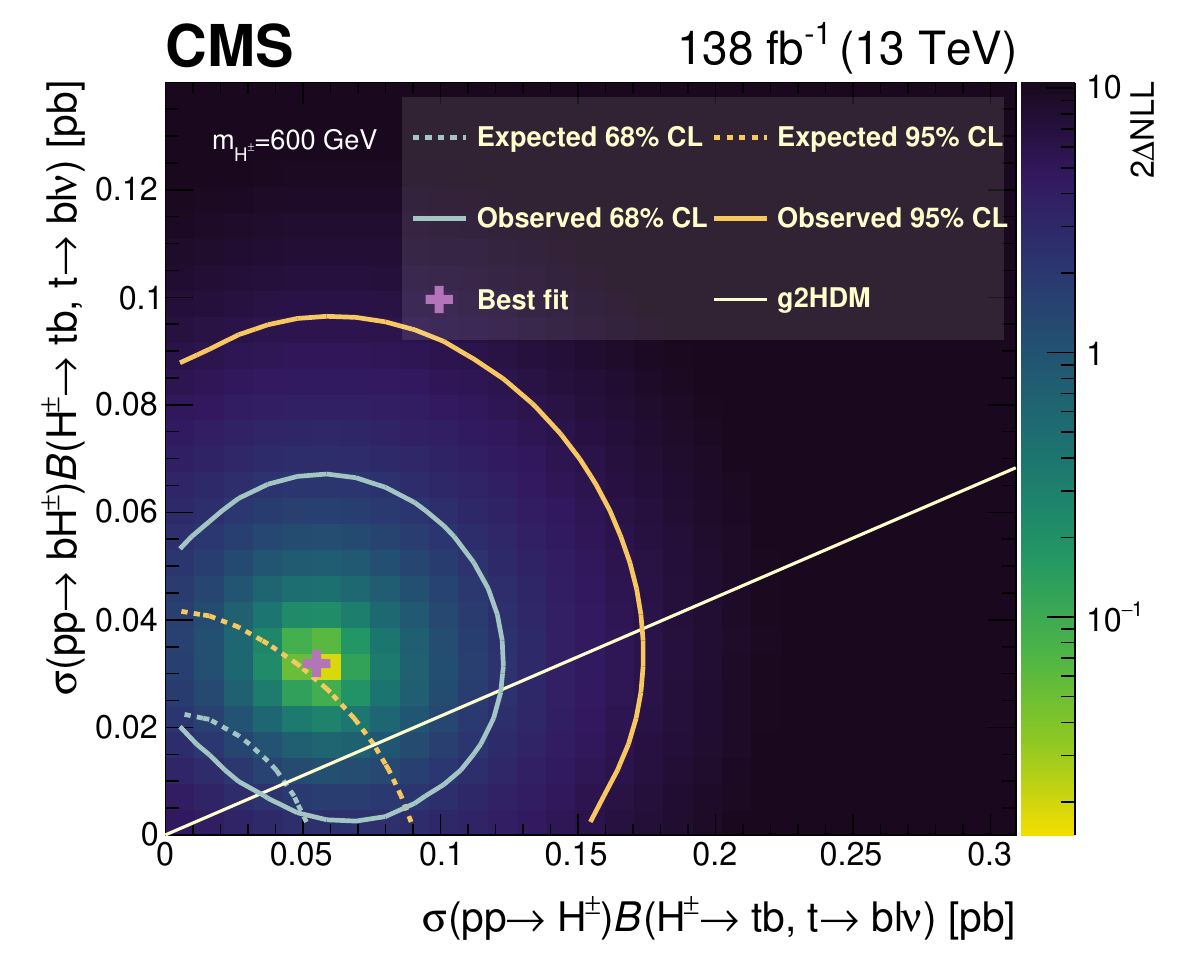}
\includegraphics[width=0.75\textwidth]{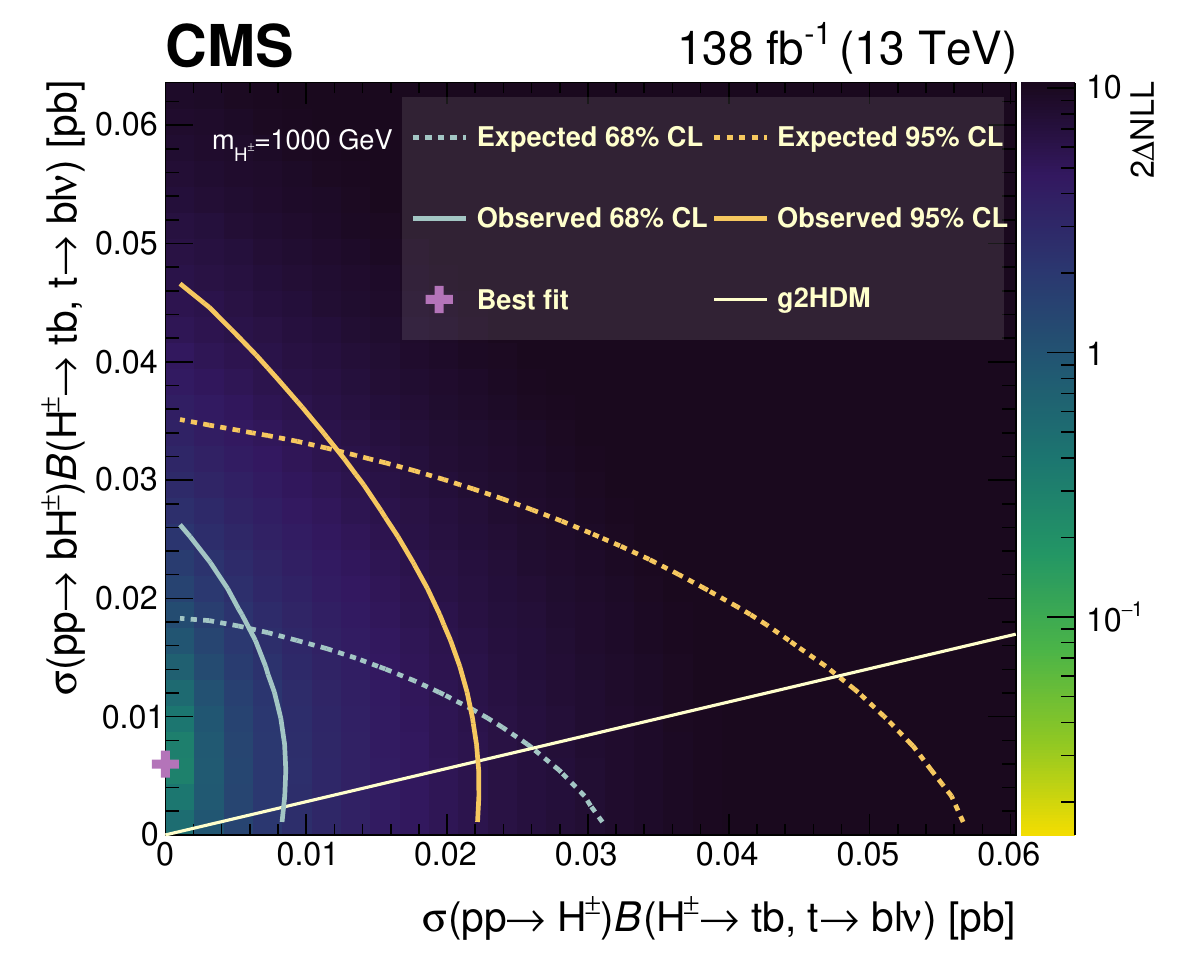}
\caption{Two times the difference of the negative log-likelihood (NLL) as a function of \crosssectionlight\brhtb and \crosssectionbH\brhtb
with the best fit point extracted from the pDNN distribution for the
\Hpm boson mass $\mHpm = 600 \GeV$ (upper), and $\mHpm = 1000 \GeV$ (lower).
The solid and dashed contours represent the observed and expected limits, respectively.
The points along the contours represent 68 and 95\% \CL regions, extracted from 
the $2\Delta{\text{NLL}}$ values of 2.30 and 5.99, respectively.
The g2HDM predictions are also shown. 
The points along the g2HDM prediction line represent different $\rhott$, $\rhotc$ coupling sets, with all other extra Yukawa couplings assumed to be zero.}
\label{fig:limit_contour_run2_mH500}
\end{figure*}

In order to provide cross section upper limits independent of the g2HDM, we separate the signal samples into two categories based on 
the number of \PQb quarks at the parton level: one with two \PQb quarks (\crosssectionlight) 
and another with three or more \PQb quarks (\crosssectionbH). 
The contribution of the processes in the two categories can vary depending on different signal hypothesis, while the ratio stays constant in the g2HDM for a fixed \mHpm. 
Using these two categories, we provide a two dimensional contour scan of two times the difference of 
negative log-likelihood ($2\Delta{\text{NLL}}$) 
for \crosssectionlight\brhtb vs \crosssectionbH\brhtb  processes 
in Fig.~\ref{fig:limit_contour_run2_mH500}, 
for $\mHpm = 600$ and 1000\GeV. 
The best fit values for $\mHpm = 600$ and 1000\GeV are $\crosssectionlight\brhtb = 0.055^{+0.068}_{-0.055}$ and $0.000^{+0.009}_{-0.000}\pb$, 
and $\crosssectionbH\brhtb = 0.032^{+0.035}_{-0.029}$ and $0.006^{+0.019}_{-0.006}\pb$, respectively, 
with the corresponding 68 and 95\% \CL contours also shown in Fig.~\ref{fig:limit_contour_run2_mH500}.

\clearpage
\newpage

\begin{table}[htbp]
\centering
\caption{Summary of the systematic uncertainty decomposition with respect to the total uncertainty in the signal strength for the mass hypothesis $\mHpm$ = 200 \GeV.}
\begin{tabular}{lcc}
Uncertainty source & Down variation [\%] & Up variation [\%]\\
\hline
Theoretical background cross sections & 50.0 & 52.6 \\
Statistical uncertainty & 39.2 & 41.8 \\
Top quark \pt reweighting & 34.8 & 34.5 \\
Heavy-flavor tagging & 32.6 & 31.8 \\
Nonprompt-lepton bkg. estimate & 30.2 & 25.9 \\
Parton shower ISR/FSR & 26.7 & 23.1 \\
$\mu_{\rm r}/\mu_{\rm f}$ & 25.8 & 23.5 \\
Jet energy scale & 23.9 & 22.0 \\
Unclustered energy & 18.2 & 20.7 \\
Pileup & 13.9 & 19.1 \\
PDF & 11.5 & 10.8 \\
Jet energy resolution & 10.7 & 10.0 \\
Integrated luminosity &  3.1 &  4.7 \\
Electron reconstruction &  4.2 &  3.0 \\
Jet pileup identification &  3.6 &  1.5 \\
Lepton identification &  3.4 &  1.6 \\
Lepton trigger efficiencies &  1.2 &  2.3 \\
L1 trigger inefficiency &  0.7 &  0.0 \\
Lepton energy scale &  0.7 &  0.0 \\
\end{tabular}
\label{tab:syst_rank_H200}
\end{table}

\begin{table}[htbp]
\centering
\caption{Summary of the systematic uncertainty decomposition with respect to the total uncertainty in the signal strength for the mass hypothesis $\mHpm$ = 600 \GeV.}
\begin{tabular}{lcc}
Uncertainty source & Down variation [\%] & Up variation [\%]\\
\hline
Heavy-flavor tagging & 56.4 & 56.7 \\
Theoretical background cross sections & 39.6 & 33.4 \\
PDF & 19.6 & 32.5 \\
Statistical uncertainty & 36.9 & 31.4 \\
Top quark \pt reweighting & 31.5 & 29.8 \\
Jet energy scale & 26.8 & 24.5 \\
$\mu_{\rm r}/\mu_{\rm f}$ & 17.3 & 24.3 \\
Unclustered energy & 22.0 & 21.2 \\
Nonprompt-lepton bkg. estimate & 22.3 & 17.7 \\
Parton shower ISR/FSR & 11.0 & 17.1 \\
Pileup &  8.4 & 12.9 \\
Jet pileup identification & 12.3 & 12.1 \\
Jet energy resolution &  9.7 & 11.0 \\
Integrated luminosity &  4.4 &  5.5 \\
L1 trigger inefficiency &  5.2 &  3.5 \\
Lepton energy scale &  4.9 &  0.6 \\
Lepton trigger efficiencies &  4.0 &  3.8 \\
Electron reconstruction &  0.0 &  0.6 \\
Lepton identification &  0.0 &  0.0 \\
\end{tabular}
\label{tab:syst_rank_H600}
\end{table}

\begin{table}[htbp]
\centering
\caption{Summary of the systematic uncertainty decomposition with respect to the total uncertainty in the signal strength for the mass hypothesis $\mHpm$ = 1000 \GeV.}
\begin{tabular}{lcc}
Uncertainty source & Down variation [\%] & Up variation [\%]\\
\hline
PDF & 54.2 & 38.5 \\
Top quark \pt reweighting & 52.8 & 56.0 \\
Jet energy scale & 39.7 & 40.7 \\
Theoretical background cross sections & 34.1 & 31.0 \\
Heavy-flavor tagging & 26.3 & 32.1 \\
Statistical uncertainty &  6.0 & 30.7 \\
Parton shower ISR/FSR & 15.8 &  9.4 \\
Nonprompt-lepton bkg. estimate & 15.7 & 18.8 \\
$\mu_{\rm r}/\mu_{\rm f}$ & 14.8 & 12.8 \\
Lepton identification &  7.9 &  3.6 \\
Electron reconstruction &  0.0 &  7.3 \\
Jet pileup identification &  1.1 &  5.8 \\
L1 trigger inefficiency &  3.8 &  5.7 \\
Pileup &  2.2 &  5.5 \\
Lepton trigger efficiencies &  3.9 &  3.8 \\
Unclustered energy &  3.3 &  0.0 \\
Jet energy resolution &  0.0 &  3.7 \\
Integrated luminosity &  2.4 &  1.4 \\
Lepton energy scale &  0.0 &  1.3 \\
\end{tabular}
\label{tab:syst_rank_H1000}
\end{table}

\clearpage
\newpage

 \section{Summary}
\label{sec:summary}
 A search for charged Higgs bosons ($\Hpm$) in proton-proton (\pp) collisions at a center-of-mass energy of 13\TeV has been presented. 
 The processes considered are \pptobHpm, with \Hpm decaying into top (\PQt) and bottom (\PQb) quarks.
We consider $\PQt \to \PQb\Pell\PGn$ decays, where $\Pell = \Pe, \PGm$. The \Hpm boson mass (\mHpm) is probed between 200\GeV and 1\TeV.
The results of the search are interpreted in the generalized 
two-Higgs-doublet model (g2HDM), assuming the real parts of extra Yukawa couplings of \Hpm to the top quark $\rhott$ 
and to the top and charm quark $\rhotc$ are both less than one, and the imaginary parts and all other extra Yukawa couplings are zero.
An excess is observed with respect to the standard model expectation with a local significance of 2.4 standard deviations for a signal with $\mHpm = 600\GeV$. 
The best fit cross section ($\sigma$) times branching fraction ($\mathcal{B}$) values corresponding to the excess at $\mHpm = 600\GeV$ are 
$\crosssectionlight\brhtb = 0.055^{+0.068}_{-0.055}\pb$  and $\crosssectionbH\brhtb = 0.032^{+0.035}_{-0.029}\pb$, where $\Pell=\Pe, \PGm$. 
Upper limits at 95\% confidence level (\CL) are set on $\sigma\mathcal{B}$ for \mHpm up to 1\TeV.
The values for $\rhotc \gtrsim$ 0.11--0.38 are excluded, depending on the \mHpm and \rhott assumptions.
The strongest exclusion is achieved near $\mHpm=400\GeV$, where values of $\rhotc>0.11$ for $\rhott=1.0$ and $\rhotc>0.12$ for $\rhott=0.3$ are excluded at 95\% \CL.
For $\rhott=1$, the analysis excludes $\rhotc>0.38$ at 95\% \CL across the mass range from 200\GeV to 1\TeV.
This is the first search for \Hpm bosons within the g2HDM framework.

\begin{acknowledgments}
We congratulate our colleagues in the CERN accelerator departments for the excellent performance of the LHC and thank the technical and administrative staffs at CERN and at other CMS institutes for their contributions to the success of the CMS effort. In addition, we gratefully acknowledge the computing centers and personnel of the Worldwide LHC Computing Grid and other centers for delivering so effectively the computing infrastructure essential to our analyses. Finally, we acknowledge the enduring support for the construction and operation of the LHC, the CMS detector, and the supporting computing infrastructure provided by the following funding agencies: SC (Armenia), BMBWF and FWF (Austria); FNRS and FWO (Belgium); CNPq, CAPES, FAPERJ, FAPERGS, and FAPESP (Brazil); MES and BNSF (Bulgaria); CERN; CAS, MoST, and NSFC (China); MINCIENCIAS (Colombia); MSES and CSF (Croatia); RIF (Cyprus); SENESCYT (Ecuador); ERC PRG, TARISTU24-TK10 and MoER TK202 (Estonia); Academy of Finland, MEC, and HIP (Finland); CEA and CNRS/IN2P3 (France); SRNSF (Georgia); BMFTR, DFG, and HGF (Germany); GSRI (Greece); NKFIH (Hungary); DAE and DST (India); IPM (Iran); SFI (Ireland); INFN (Italy); MSIT and NRF (Republic of Korea); MES (Latvia); LMTLT (Lithuania); MOE and UM (Malaysia); BUAP, CINVESTAV, CONACYT, LNS, SEP, and UASLP-FAI (Mexico); MOS (Montenegro); MBIE (New Zealand); PAEC (Pakistan); MES, NSC, and NAWA (Poland); FCT (Portugal); MESTD (Serbia); MICIU/AEI and PCTI (Spain); MOSTR (Sri Lanka); Swiss Funding Agencies (Switzerland); MST (Taipei); MHESI (Thailand); TUBITAK and TENMAK (T\"{u}rkiye); NASU (Ukraine); STFC (United Kingdom); DOE and NSF (USA).

\hyphenation{Rachada-pisek} Individuals have received support from the Marie-Curie program and the European Research Council and Horizon 2020 Grant, contract Nos.\ 675440, 724704, 752730, 758316, 765710, 824093, 101115353, 101002207, 101001205, and COST Action CA16108 (European Union); the Leventis Foundation; the Alfred P.\ Sloan Foundation; the Alexander von Humboldt Foundation; the Science Committee, project no. 22rl-037 (Armenia); the Fonds pour la Formation \`a la Recherche dans l'Industrie et dans l'Agriculture (FRIA) and Fonds voor Wetenschappelijk Onderzoek contract No. 1228724N (Belgium); the Beijing Municipal Science \& Technology Commission, No. Z191100007219010, the Fundamental Research Funds for the Central Universities, the Ministry of Science and Technology of China under Grant No. 2023YFA1605804, the Natural Science Foundation of China under Grant No. 12061141002, 12535004, and USTC Research Funds of the Double First-Class Initiative No.\ YD2030002017 (China); the Ministry of Education, Youth and Sports (MEYS) of the Czech Republic; the Shota Rustaveli National Science Foundation, grant FR-22-985 (Georgia); the Deutsche Forschungsgemeinschaft (DFG), among others, under Germany's Excellence Strategy -- EXC 2121 ``Quantum Universe" -- 390833306, and under project number 400140256 - GRK2497; the Hellenic Foundation for Research and Innovation (HFRI), Project Number 2288 (Greece); the Hungarian Academy of Sciences, the New National Excellence Program - \'UNKP, the NKFIH research grants K 131991, K 133046, K 138136, K 143460, K 143477, K 146913, K 146914, K 147048, 2020-2.2.1-ED-2021-00181, TKP2021-NKTA-64, and 2025-1.1.5-NEMZ\_KI-2025-00004 (Hungary); the Council of Science and Industrial Research, India; ICSC -- National Research Center for High Performance Computing, Big Data and Quantum Computing, FAIR -- Future Artificial Intelligence Research, and CUP I53D23001070006 (Mission 4 Component 1), funded by the NextGenerationEU program (Italy); the Latvian Council of Science; the Ministry of Education and Science, project no. 2022/WK/14, and the National Science Center, contracts Opus 2021/41/B/ST2/01369, 2021/43/B/ST2/01552, 2023/49/B/ST2/03273, and the NAWA contract BPN/PPO/2021/1/00011 (Poland); the Funda\c{c}\~ao para a Ci\^encia e a Tecnologia, grant CEECIND/01334/2018 (Portugal); the National Priorities Research Program by Qatar National Research Fund; MICIU/AEI/10.13039/501100011033, ERDF/EU, ``European Union NextGenerationEU/PRTR", and Programa Severo Ochoa del Principado de Asturias (Spain); the Chulalongkorn Academic into Its 2nd Century Project Advancement Project, the National Science, Research and Innovation Fund program IND\_FF\_68\_369\_2300\_097, and the Program Management Unit for Human Resources \& Institutional Development, Research and Innovation, grant B39G680009 (Thailand); the Kavli Foundation; the Nvidia Corporation; the SuperMicro Corporation; the Welch Foundation, contract C-1845; and the Weston Havens Foundation (USA).
\end{acknowledgments}\section*{Data availability} Release and preservation of data used by the CMS Collaboration as the basis for publications is guided by the  \href{https://doi.org/10.7483/OPENDATA.CMS.1BNU.8V1W}{CMS data preservation, re-use and open access policy}.

\bibliography{auto_generated} 
\cleardoublepage \appendix\section{The CMS Collaboration \label{app:collab}}\begin{sloppypar}\hyphenpenalty=5000\widowpenalty=500\clubpenalty=5000\input{B2G-24-008-public-authorlist.tex}\end{sloppypar}
\end{document}

%% file: B2G-24-008-public-authorlist.tex
\cmsinstitute{Yerevan Physics Institute, Yerevan, Armenia}
{\tolerance=6000
A.~Hayrapetyan, V.~Makarenko\cmsorcid{0000-0002-8406-8605}, A.~Tumasyan\cmsAuthorMark{1}\cmsorcid{0009-0000-0684-6742}
\par}
\cmsinstitute{Institut f\"{u}r Hochenergiephysik, Vienna, Austria}
{\tolerance=6000
W.~Adam\cmsorcid{0000-0001-9099-4341}, L.~Benato\cmsorcid{0000-0001-5135-7489}, T.~Bergauer\cmsorcid{0000-0002-5786-0293}, M.~Dragicevic\cmsorcid{0000-0003-1967-6783}, C.~Giordano, P.S.~Hussain\cmsorcid{0000-0002-4825-5278}, M.~Jeitler\cmsAuthorMark{2}\cmsorcid{0000-0002-5141-9560}, N.~Krammer\cmsorcid{0000-0002-0548-0985}, A.~Li\cmsorcid{0000-0002-4547-116X}, D.~Liko\cmsorcid{0000-0002-3380-473X}, M.~Matthewman, J.~Schieck\cmsAuthorMark{2}\cmsorcid{0000-0002-1058-8093}, R.~Sch\"{o}fbeck\cmsAuthorMark{2}\cmsorcid{0000-0002-2332-8784}, D.~Schwarz\cmsorcid{0000-0002-3821-7331}, M.~Shooshtari\cmsorcid{0009-0004-8882-4887}, M.~Sonawane\cmsorcid{0000-0003-0510-7010}, W.~Waltenberger\cmsorcid{0000-0002-6215-7228}, C.-E.~Wulz\cmsAuthorMark{2}\cmsorcid{0000-0001-9226-5812}
% ...
\par}
\vskip\cmsinstskip
\dag:~Deceased\\
$^{1}$Also at Yerevan State University, Yerevan, Armenia\\
$^{2}$Also at TU Wien, Vienna, Austria\\
% ...
